\documentclass{aa}

\usepackage{natbib}
\bibpunct{(}{)}{;}{a}{}{,} 
\usepackage{amssymb}
\usepackage{url}
\usepackage{hyperref}
\usepackage{graphicx}
\usepackage{longtable}
\usepackage[usenames,dvipsnames]{color}
\usepackage{booktabs}
\usepackage{lscape}
\newcommand{\urltt}[1]{\texttt{#1}}

\newcommand{\msun}{\mbox{$M_\odot$}}
\newcommand{\mstar}{\mbox{$M_*$}}

\newcommand{\msunyr}{\mbox{\msun\ yr$^{-1}$}}
\newcommand{\hi}{\ion{H}{i}}
\newcommand{\mhi}{\mbox{$M_{\rm H{\sc I}}$}}

\newcommand{\lp}{\mbox{$L'$}}
\newcommand{\kms}{\mbox{km\,s$^{-1}$}}
\newcommand{\Kkmspc}{\mbox{K\,km\,s$^{-1}$\,pc$^2$}}
\newcommand{\htwo}{\mbox{H$_2$}}
\newcommand{\mhtwo}{\mbox{$M_{\rm H_2}$}}
\newcommand{\metoh}{12+\log(\mbox{O}/\mbox{H})}

\urldef{\hyperleda}\url{http://leda.univ-lyon1.fr/ledacat.cgi?NGC%203278}

\newcommand{\smgs}{submm galaxies}
\newcommand{\lpcoone}{\mbox{$L'_{\rm 1-0}$}}
\newcommand{\lpcotwo}{\mbox{$L'_{\rm 2-1}$}}
\newcommand{\lpcothree}{\mbox{$L'_{\rm 3-2}$}}
\newcommand{\lpcofour}{\mbox{$L'_{\rm 4-3}$}}

\begin{document}

 \title{
 Molecular gas masses of gamma-ray burst host galaxies}
 
\titlerunning{Molecular gas masses of gamma-ray burst host galaxies}
\authorrunning{Micha{\l}owski et al.}

\author{Micha{\l}~J.~Micha{\l}owski\inst{\ref{inst:poz},\ref{inst:roe}
}
\and
A.~Karska\inst{\ref{inst:torun}}
\and
J.~R.~Rizzo\inst{\ref{inst:csic}}
\and
M.~Baes\inst{\ref{inst:gent}}
\and
A.~J.~Castro-Tirado\inst{\ref{inst:ant}}
\and
J.~Hjorth\inst{\ref{inst:dark}}
\and
L.~K.~Hunt\inst{\ref{inst:hunt}}
\and
P.~Kamphuis\inst{\ref{inst:rub},\ref{inst:gmrt}}
\and
M.~P.~Koprowski\inst{\ref{inst:herts}}
\and
M.~R.~Krumholz\inst{\ref{inst:krum}}    
\and
D.~Malesani\inst{\ref{inst:dark}}       
\and
A.~Nicuesa Guelbenzu\inst{\ref{inst:taut}}
\and
J.~Rasmussen\inst{\ref{inst:dark},\ref{inst:dtu}}
\and
A.~Rossi\inst{\ref{inst:pal}}                           
\and 
P.~Schady\inst{\ref{inst:mpe}}                          
\and
J.~Sollerman\inst{\ref{inst:soll}}                              
\and
P.~van der Werf\inst{\ref{inst:vdw}}                    
        }

\institute{
Astronomical Observatory Institute, Faculty of Physics, Adam Mickiewicz University, ul.~S{\l}oneczna 36, 60-286 Pozna{\'n}, Poland, {\tt mj.michalowski@gmail.com}\label{inst:poz}
\and
SUPA\thanks{Scottish Universities Physics Alliance}, Institute for Astronomy, University of Edinburgh, Royal Observatory, Blakford Hill, Edinburgh, EH9 3HJ, UK
\label{inst:roe}
\and
Centre for Astronomy, Faculty of Physics, Astronomy and Informatics, Nicolaus Copernicus University, Grudzi\c{a}dzka 5, 87-100 Toru\'{n}, Poland \label{inst:torun}
\and
Centro de Astrobiolog\'{\i}a (INTA-CSIC), Ctra. M-108, km.~4, E-28850 Torrej\'on de Ardoz, Madrid, Spain \label{inst:csic}
\and           
Sterrenkundig Observatorium, Universiteit Gent, Krijgslaan 281-S9, 9000, Gent, Belgium  \label{inst:gent}
\and
Instituto de Astrof\' isica de Andaluc\' ia (IAA-CSIC), Glorieta de la Astronom\' ia s/n, E-18008, Granada, Spain \label{inst:ant}
\and
Dark Cosmology Centre, Niels Bohr Institute, University of Copenhagen, Juliane Maries Vej 30, DK-2100 Copenhagen \O, Denmark  \label{inst:dark}
\and
INAF-Osservatorio Astrofisico di Arcetri, Largo E. Fermi 5, I-50125 Firenze, Italy \label{inst:hunt}
\and
Astronomisches Institut der Ruhr-Universit\"{a}t Bochum (AIRUB), Universit\"{a}tsstrasse 150, 44801 Bochum, Germany\label{inst:rub}
\and
National Centre for Radio Astrophysics, TIFR, Ganeshkhind, Pune 411007, India \label{inst:gmrt}
\and
Centre for Astrophysics Research, University of Hertfordshire, College Lane, Hatfield AL10 9AB, UK \label{inst:herts}
\and
Research School of Astronomy and Astrophysics, Australian National University, Canberra, ACT, Australia \label{inst:krum}
\and
Th\"uringer Landessternwarte Tautenburg, Sternwarte 5, D-07778 Tautenburg, Germany \label{inst:taut}
\and
Technical University of Denmark, Department of Physics, Fysikvej, building 309, DK-2800 Kgs. Lyngby, Denmark \label{inst:dtu}
\and
INAF-OAS, via Piero Gobetti 93/3 - 40129 Bologna - Italy
\label{inst:pal}
\and
Max-Planck-Institut f\"{u}r Extraterrestrische Physik, Giessenbachstra{\ss}e, D-85748 Garching bei M\"{u}nchen, Germany \label{inst:mpe}
\and
The Oskar Klein Centre, Department of Astronomy, AlbaNova, Stockholm University, 106 91 Stockholm, Sweden\label{inst:soll} 
\and
Leiden Observatory, Leiden University, P.O. Box 9513, NL-2300 RA Leiden, The Netherlands \label{inst:vdw}
}

\abstract{%
}
{The objectives of this paper are to analyse molecular gas properties of the first substantial sample of GRB hosts and test whether they are deficient in molecular gas.
}
{We obtained CO(2-1) observations of seven GRB hosts with the APEX 
and IRAM 30m telescopes.
We analysed these data together with all other hosts with previous CO observations.
}
{We obtained detections for 3 GRB hosts (980425, 080207, and 111005A) and upper limits for the remaining 4 (031203, 060505, 060814, and 100316D). 
In our entire sample of 12 CO-observed GRB hosts, 3 are clearly deficient in molecular gas, even taking into account their metallicity (980425, 060814, and 080517).
Four others are close to the best-fit line for other star-forming galaxies on the SFR-{\mhtwo} plot (051022, 060505, 080207, and 100316D). 
One host is clearly molecule rich (111005A). 
Finally, the data for 4 GRB hosts are not deep enough to judge whether they are molecule deficient (000418, 030329, 031203, and 090423).
The median value of the molecular gas depletion time, $\mhtwo/\mbox{SFR}$, of GRB hosts is $\sim0.3$\,dex below that of other star-forming galaxies, but this result has low statistical significance. A Kolmogorov-Smirnov test performed on $\mhtwo/\mbox{SFR}$ shows an only $\sim2\sigma$ difference between GRB hosts and other galaxies. This difference can partly be explained by metallicity effects, since the significance decreases to $\sim1\sigma$ for $\mhtwo/\mbox{SFR}$ versus~metallicity.
}
{We found that any molecular gas deficiency of GRB hosts has low statistical significance and that it can be attributed to their lower metallicities; and thus the sample of GRB hosts has molecular properties that are consistent with those of other galaxies, and they can be treated as representative star-forming galaxies.
However, the molecular gas deficiency can be strong for GRB hosts if they exhibit higher excitations and/or a lower CO-to-{\htwo} conversion factor than we assume, which would lead to lower molecular gas masses than we derive.
Given the concentration of atomic gas recently found close to GRB and supernova sites, indicating recent gas inflow, our results about the weak molecular deficiency imply that such an inflow does not enhance the SFRs significantly, or that atomic gas converts efficiently into the molecular phase, which fuels star formation. 
}

\maketitle

\section{Introduction}
\label{sec:intro}

Long gamma-ray bursts (GRBs) have long been confirmed to be the endpoints of lives of very massive stars \citep[e.g.][]{hjorthnature,stanek,hjorthsn}. Most of the tracers of the star formation rate (SFR) of galaxies are connected with emission from massive stars \citep[e.g.][]{kennicutt}, so that GRBs were also used to measure the star formation history of the Universe
\citep{yuksel08,kistler09,butler10,elliott12,robertson12,perley16,perley16b}.
This approach is valid if GRB hosts are representative star-forming galaxies at a given redshift \citep{michalowski12grb,hunt14,schady14, greiner15,kohn15}, or if biases are known and can be corrected for \citep{perley13,perley15,perley16,perley16b,boissier13,vergani15,schulze15,greiner16}. Gas is the fuel of star formation, so one of the important aspects of this issue is whether GRB hosts exhibit normal gas properties with respect to other star-forming galaxies.

The information about gas properties of GRB hosts is scarce. \citet{michalowski15hi} and \citet{arabsalmani15b} provided the only measurements so far of the atomic gas properties of five such galaxies. This led to a suggestion that GRB hosts have experienced  recent inflows of atomic gas. A resulting possibility of using GRBs to select galaxies for the study of gas accretion is important because 
the rate of the gas accretion onto galaxies 
is surprisingly constant since $z\sim5$, which is at odds with the significantly changing SFR volume density of the Universe \citep{spring17}.
Moreover, a fraction of star formation in GRB hosts may be directly fuelled by atomic gas \citep{michalowski15hi,michalowski16}. The existence of this process is controversial, but it has been predicted theoretically \citep{glover12,krumholz12,hu16,elmegreen18} and is supported by some observations \citep{bigiel10,fumagalli08,elmegreen16}.

Clearly, most of the star formation in the Universe is fuelled by molecular gas \citep{fumagalli09,carilli13,rafelski16}. There were several unsuccessful searches of CO lines for GRB hosts \citep{kohno05,endo07,hatsukade07,hatsukade11b,stanway11} and only four detections so far, for the hosts of GRB 980425 \citep{michalowski16},
051022  \citep[][]{hatsukade14},
080517 \citep[][]{stanway15},
and 080207 \citep{arabsalmani18}.
These studies resulted in mixed conclusions on whether GRB hosts are deficient in molecular gas with respect to the SFR-{\mhtwo} correlation of other star-forming galaxies.

Hence, the objectives of this paper are {\it i)} to analyse molecular gas properties of the first substantial sample of GRB hosts;  and {\it ii)} to test whether these hosts are deficient in molecular gas.
For this, we combined existing literature data with new observations using the APEX and IRAM 30m telescopes.

We use a cosmological model with $H_0=70$ km s$^{-1}$ Mpc$^{-1}$,  $\Omega_\Lambda=0.7$, and $\Omega_m=0.3$.
We also assume the 
\citet{chabrier03} 
initial mass function (IMF), to which all star formation rates (SFRs) and stellar masses were converted (by dividing by 1.8) if given originally assuming the \citet{salpeter} IMF.

\section{Target selection and data}
\label{sec:data}

\newlength{\mycolwi}
\setlength{\mycolwi}{0.24\textwidth}

\begin{figure*}
\begin{center}
\begin{tabular}{cccc}
\includegraphics[width= \mycolwi]{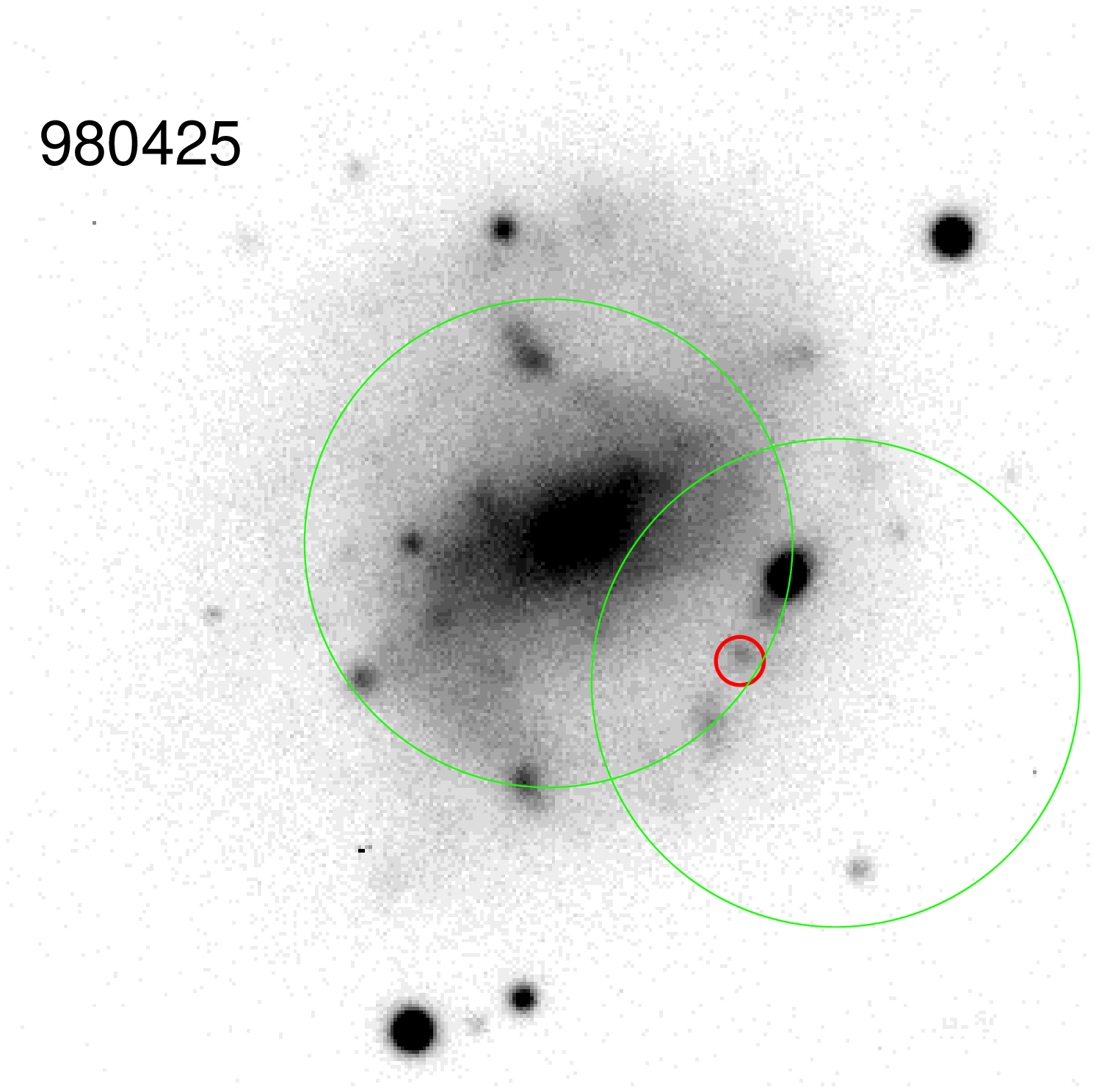} &
        \includegraphics[width= \mycolwi]{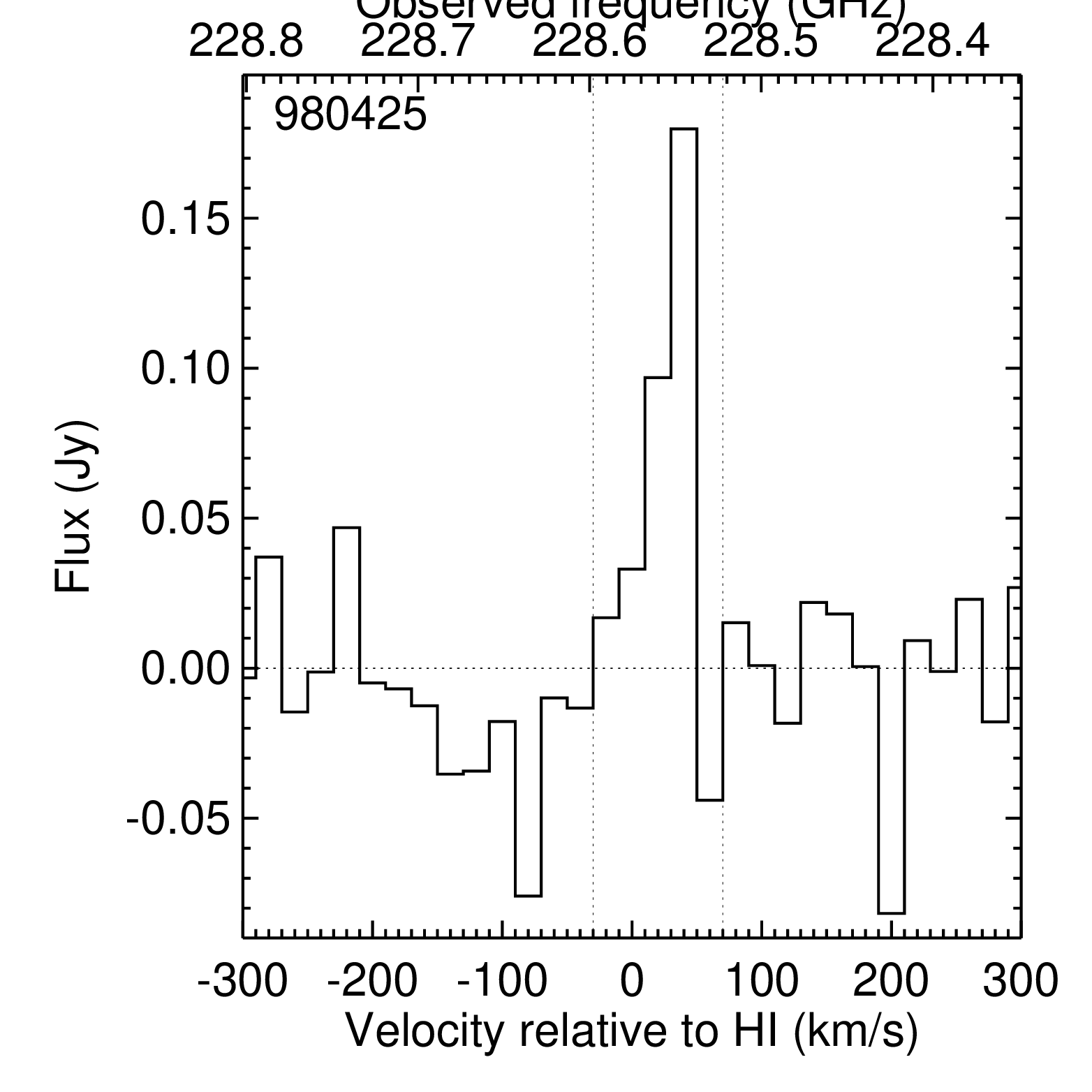} &
        \includegraphics[width= \mycolwi]{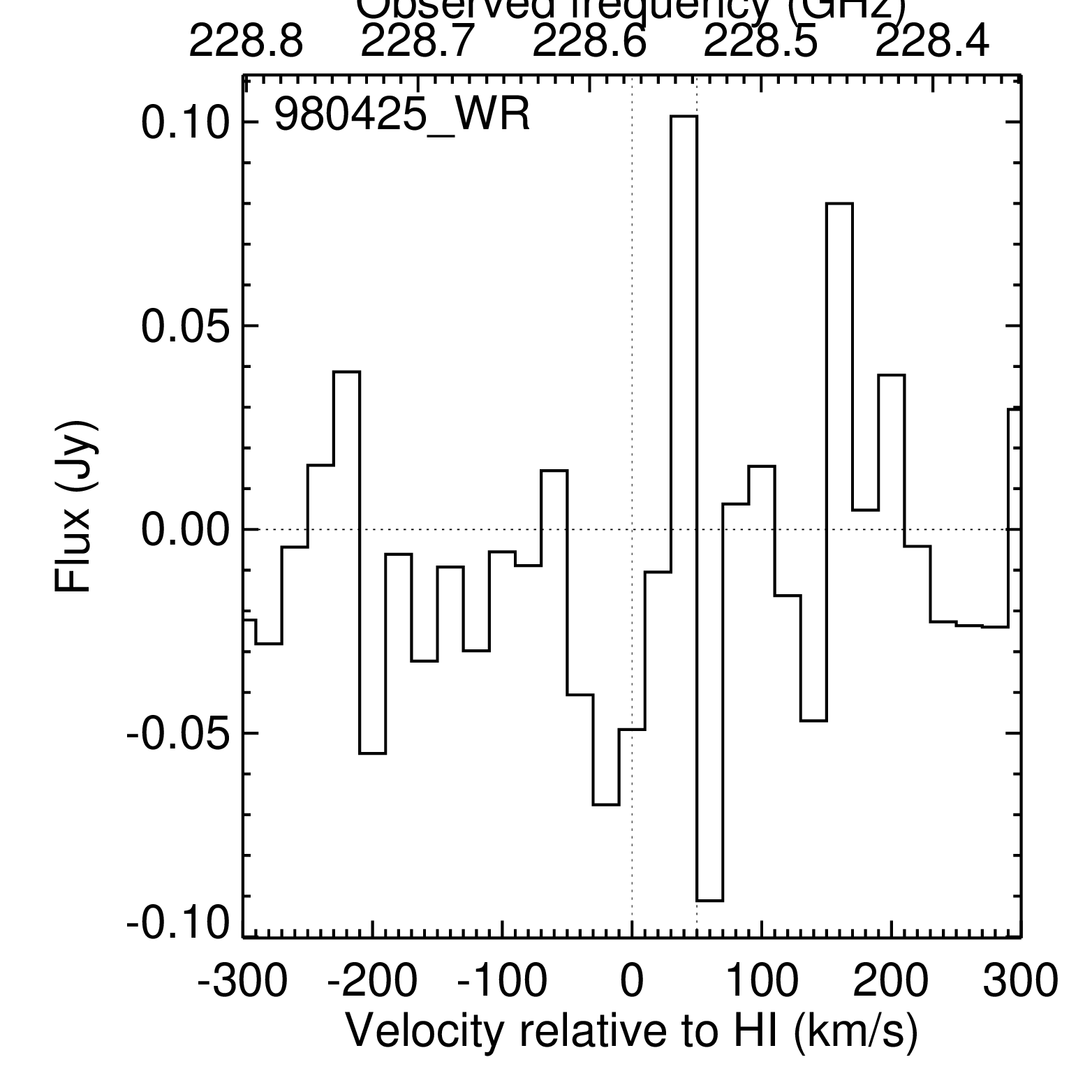}\\
\includegraphics[width= \mycolwi]{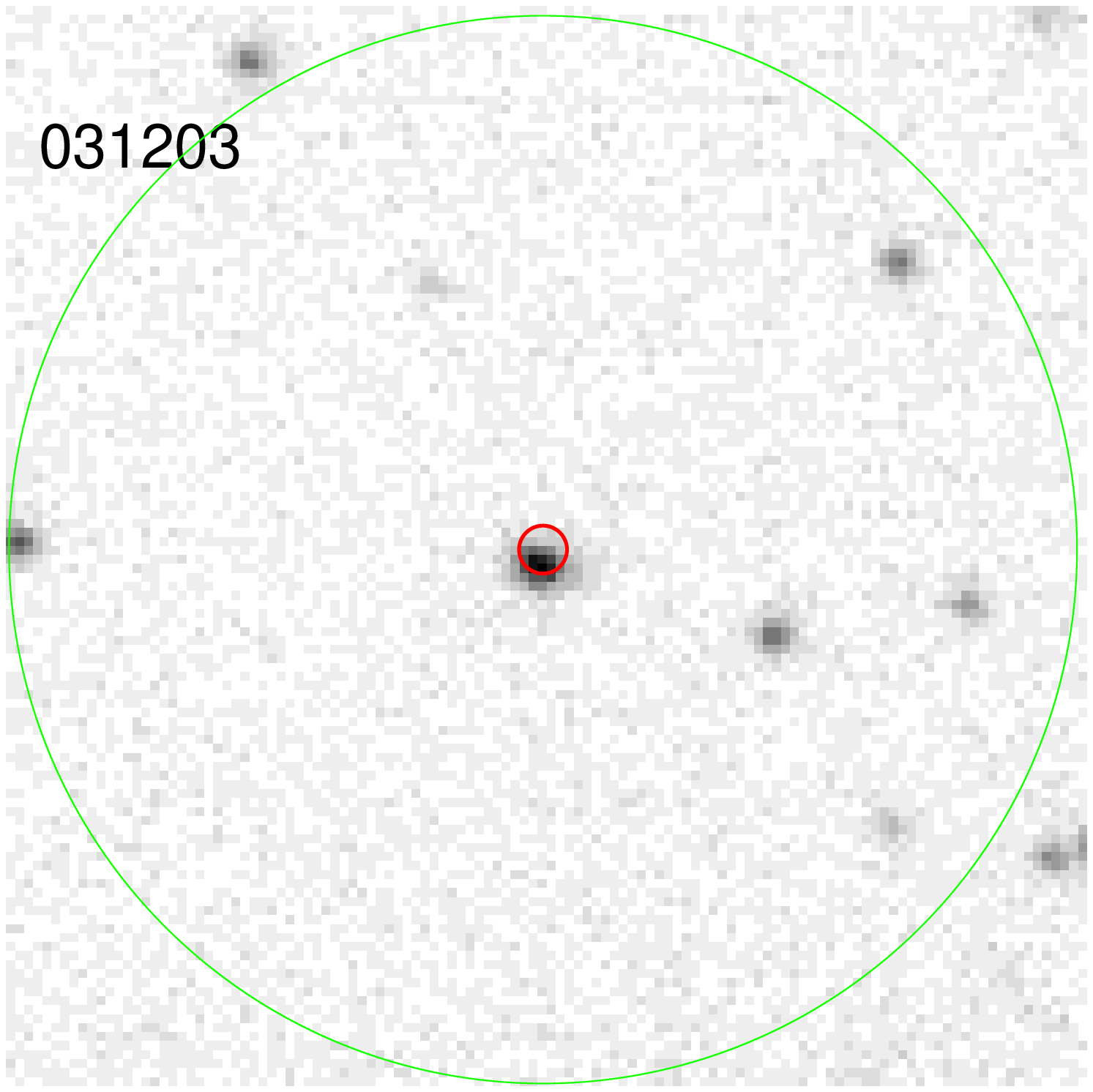} &
        \includegraphics[width= \mycolwi]{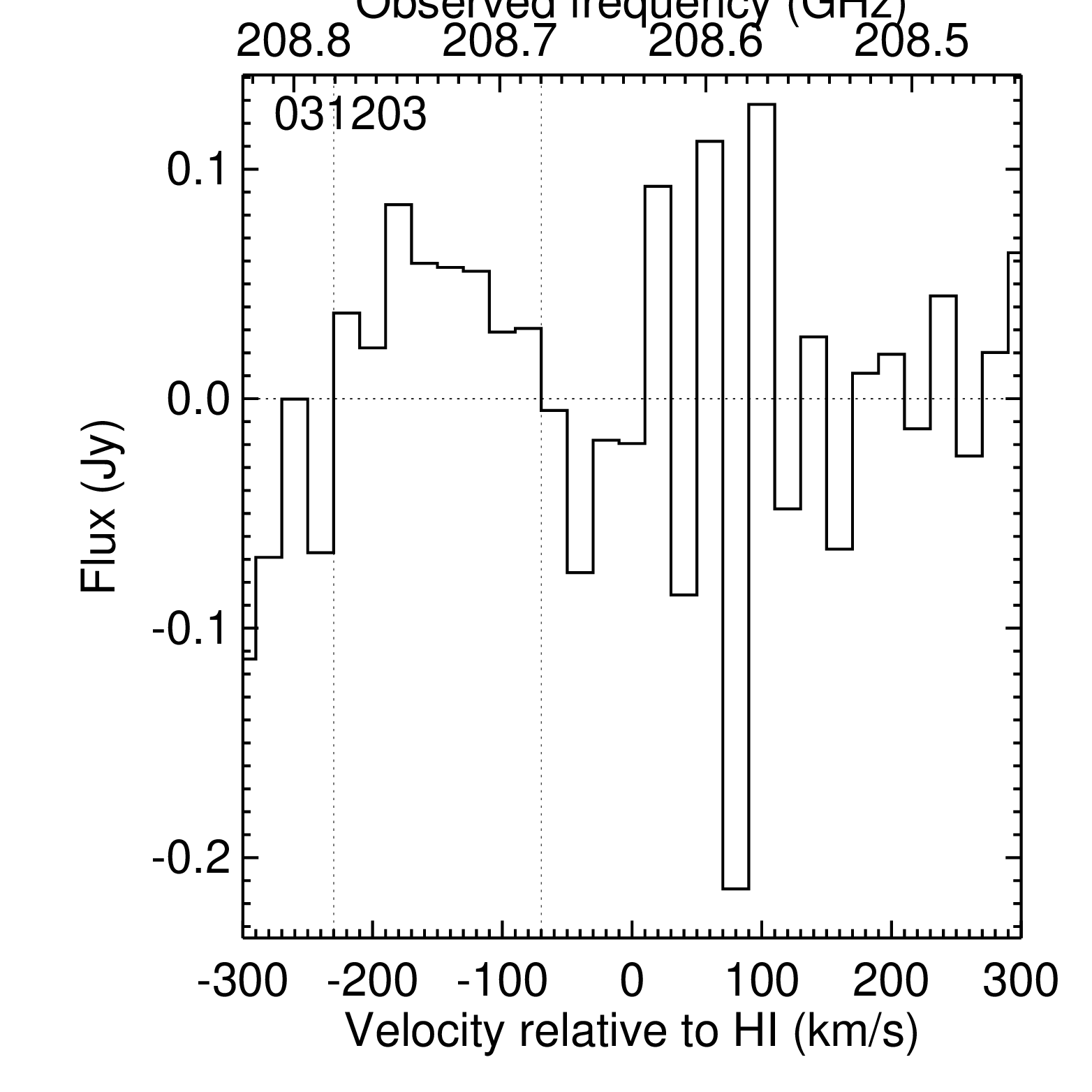} & 
\includegraphics[width= \mycolwi]{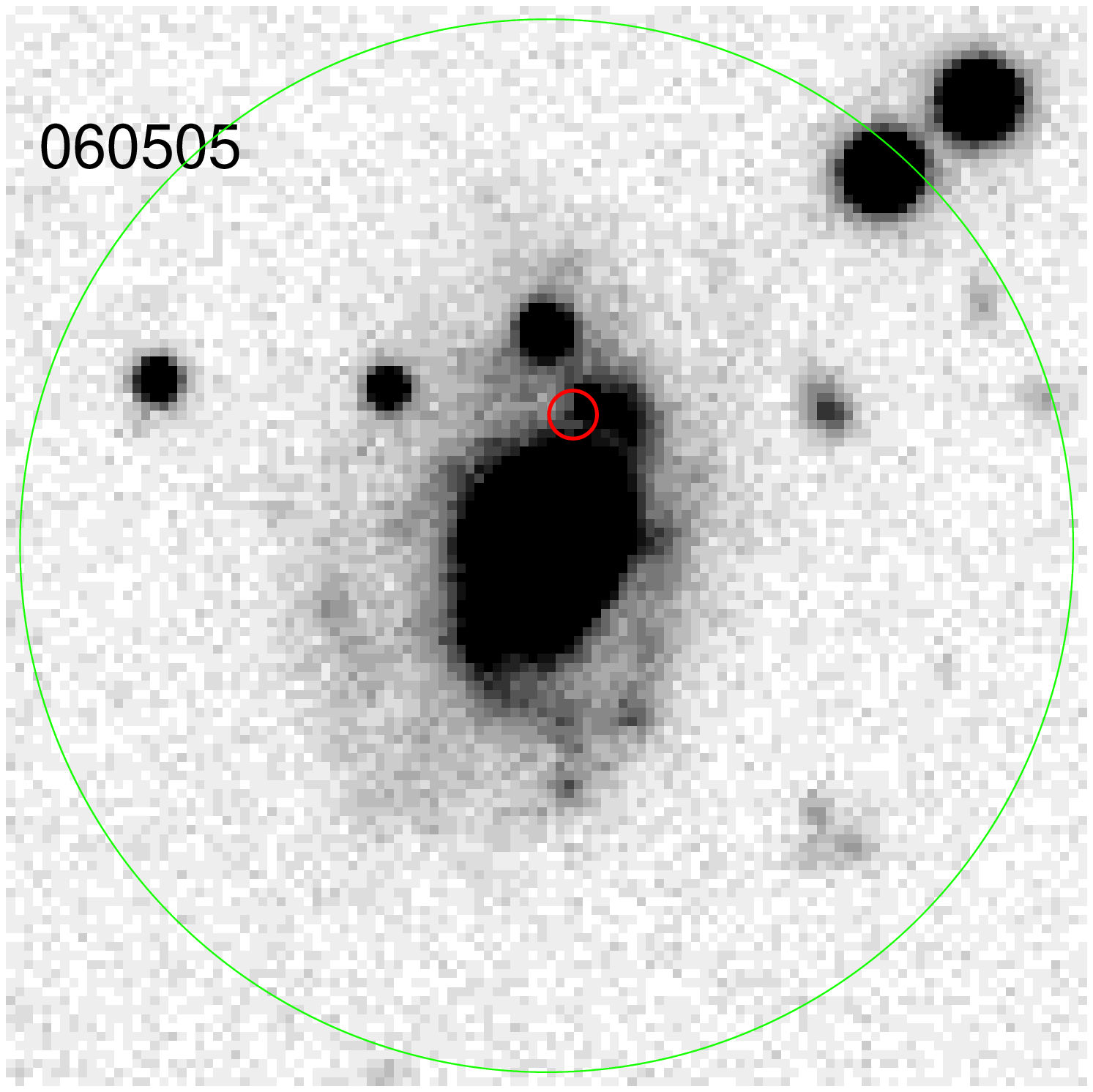} &
        \includegraphics[width= \mycolwi]{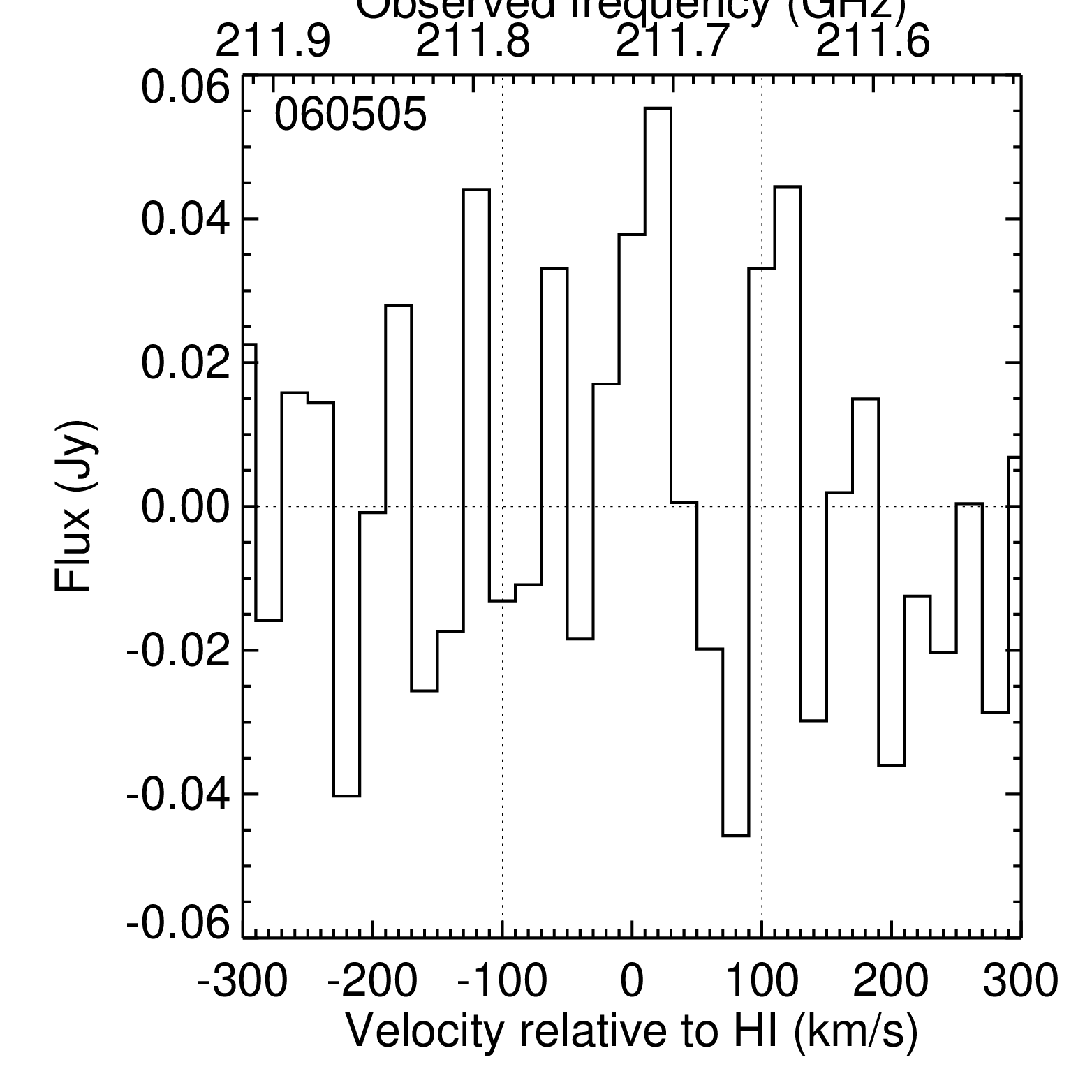} \\
\includegraphics[width= \mycolwi]{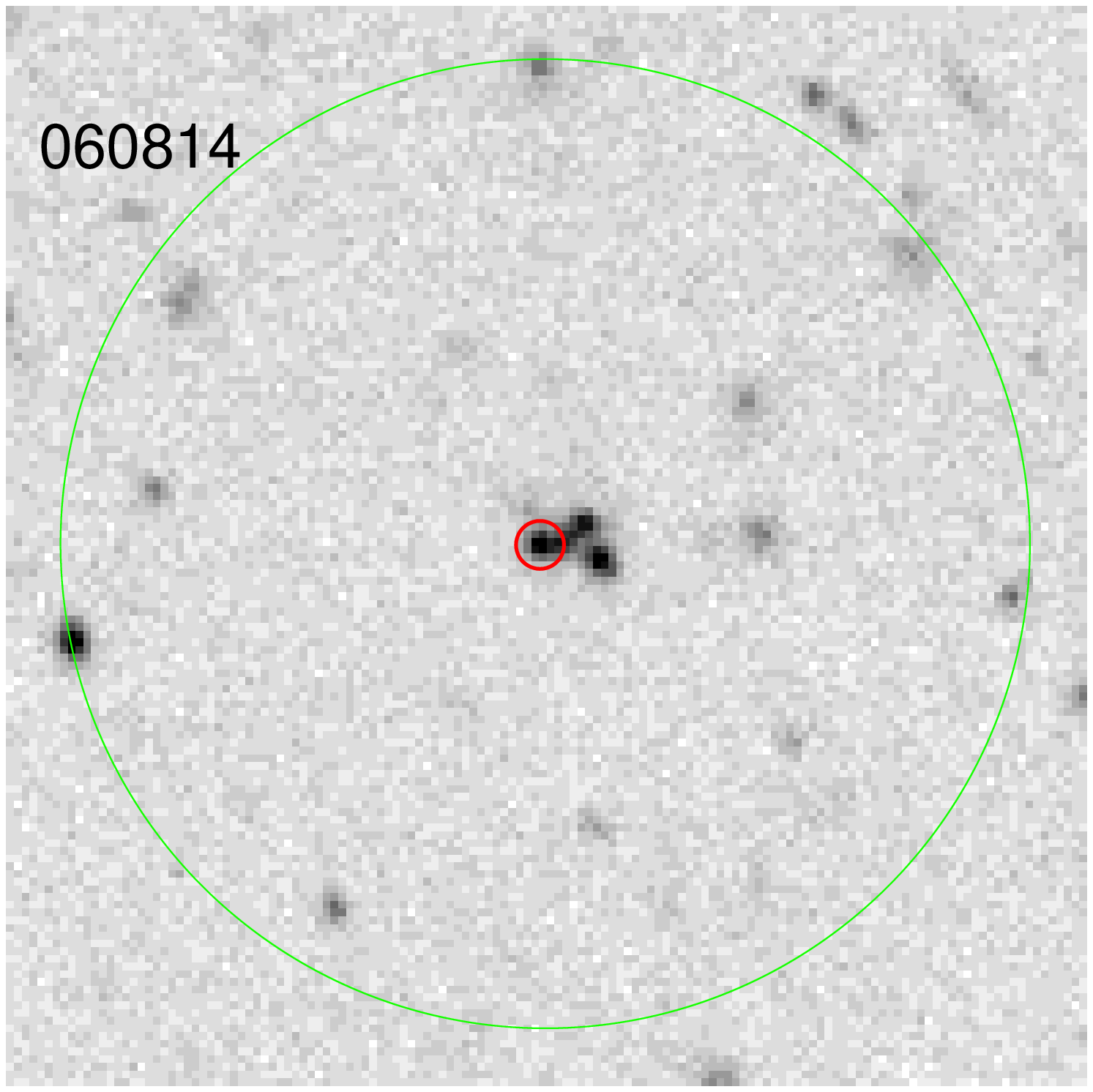} &
        \includegraphics[width= \mycolwi]{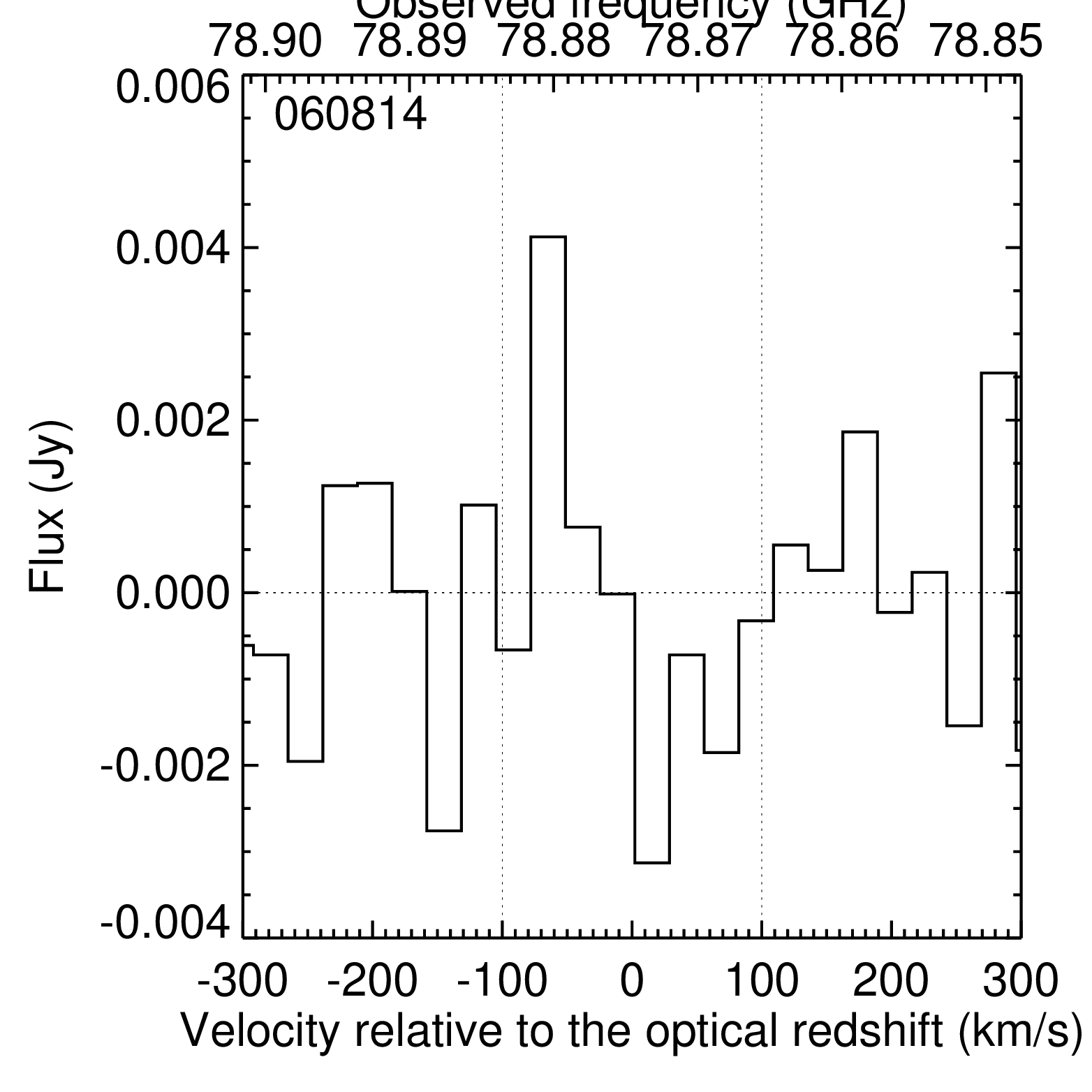} & 
        

\includegraphics[width= \mycolwi]{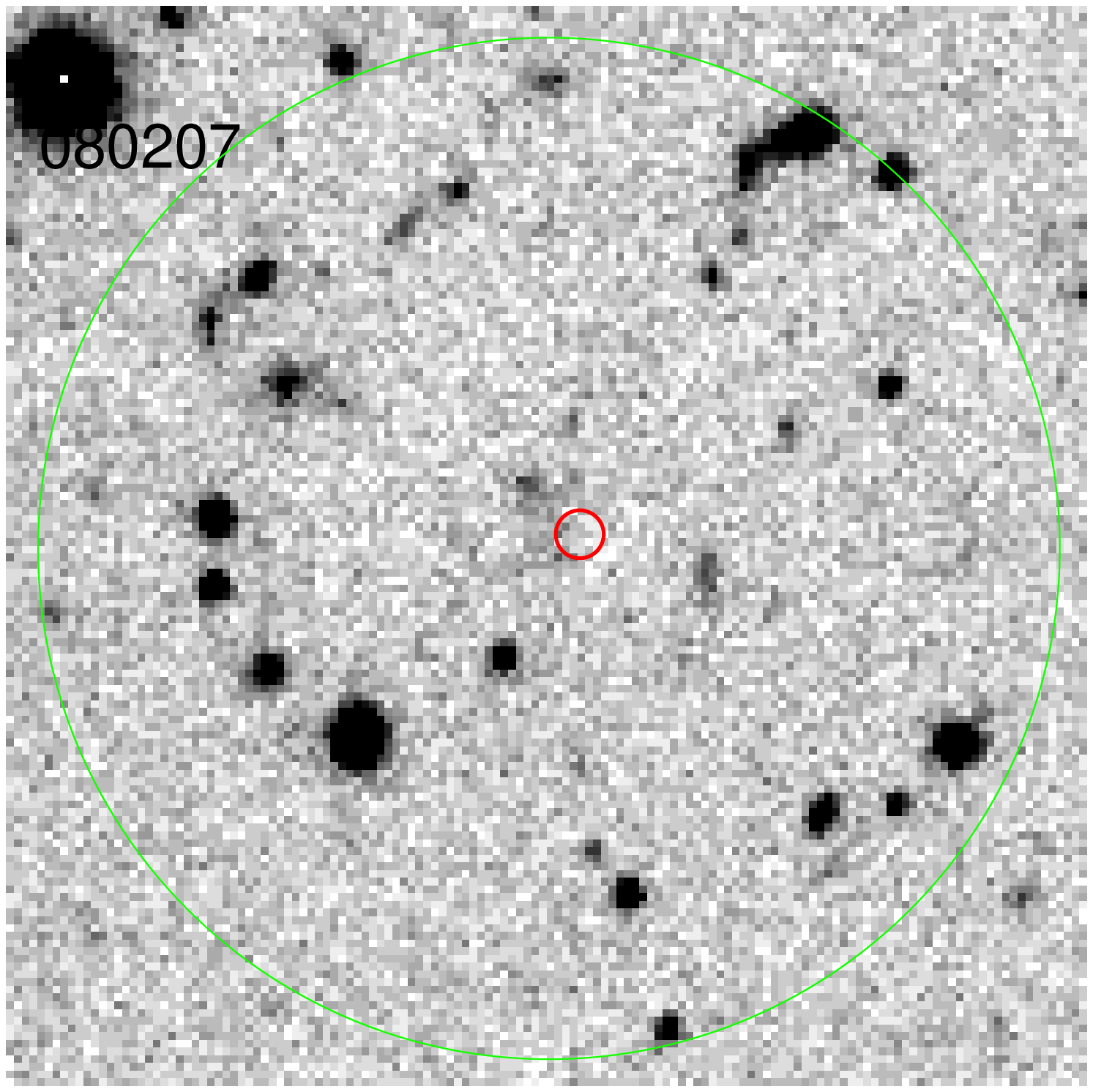} &
        \includegraphics[width= \mycolwi]{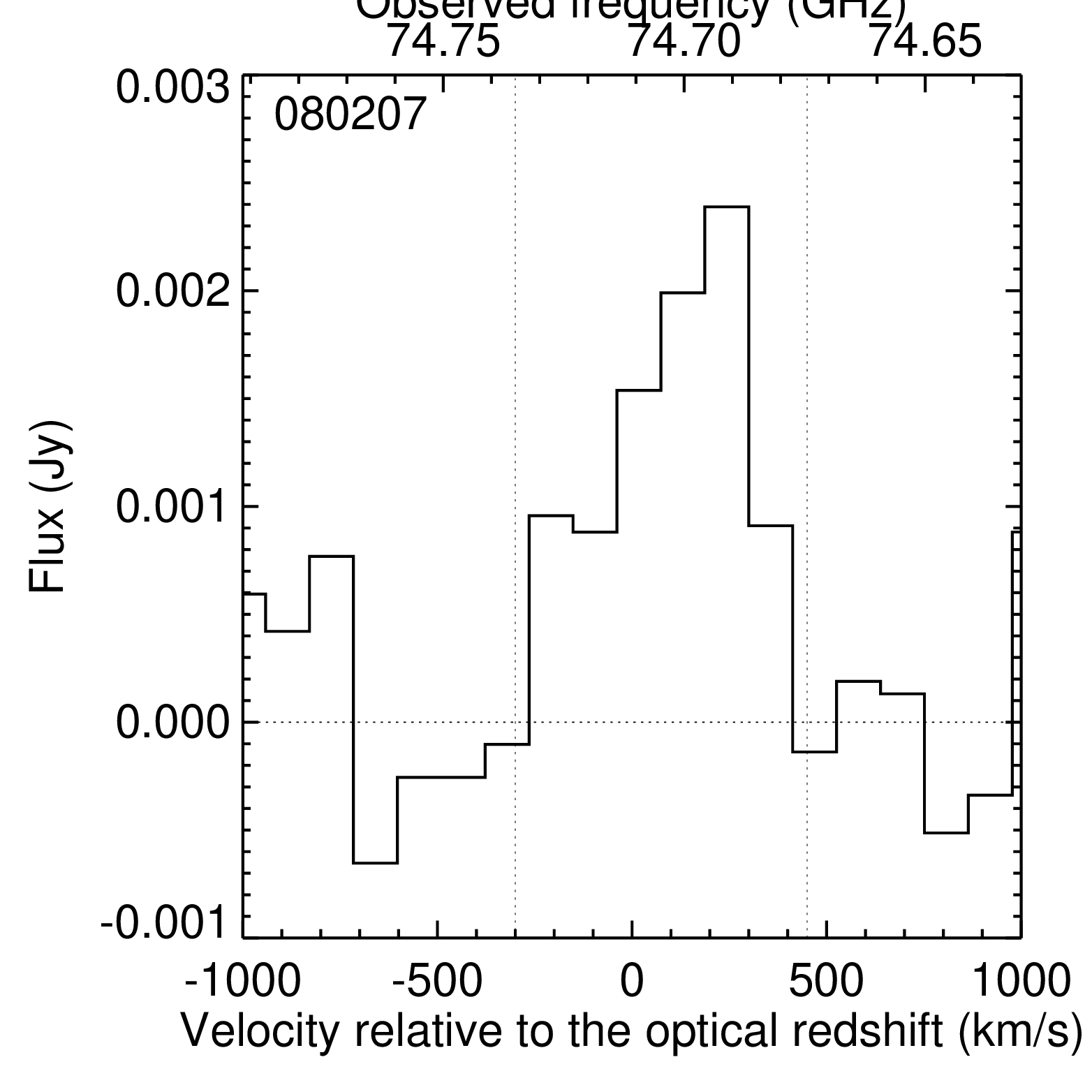} \\
\includegraphics[width= \mycolwi]{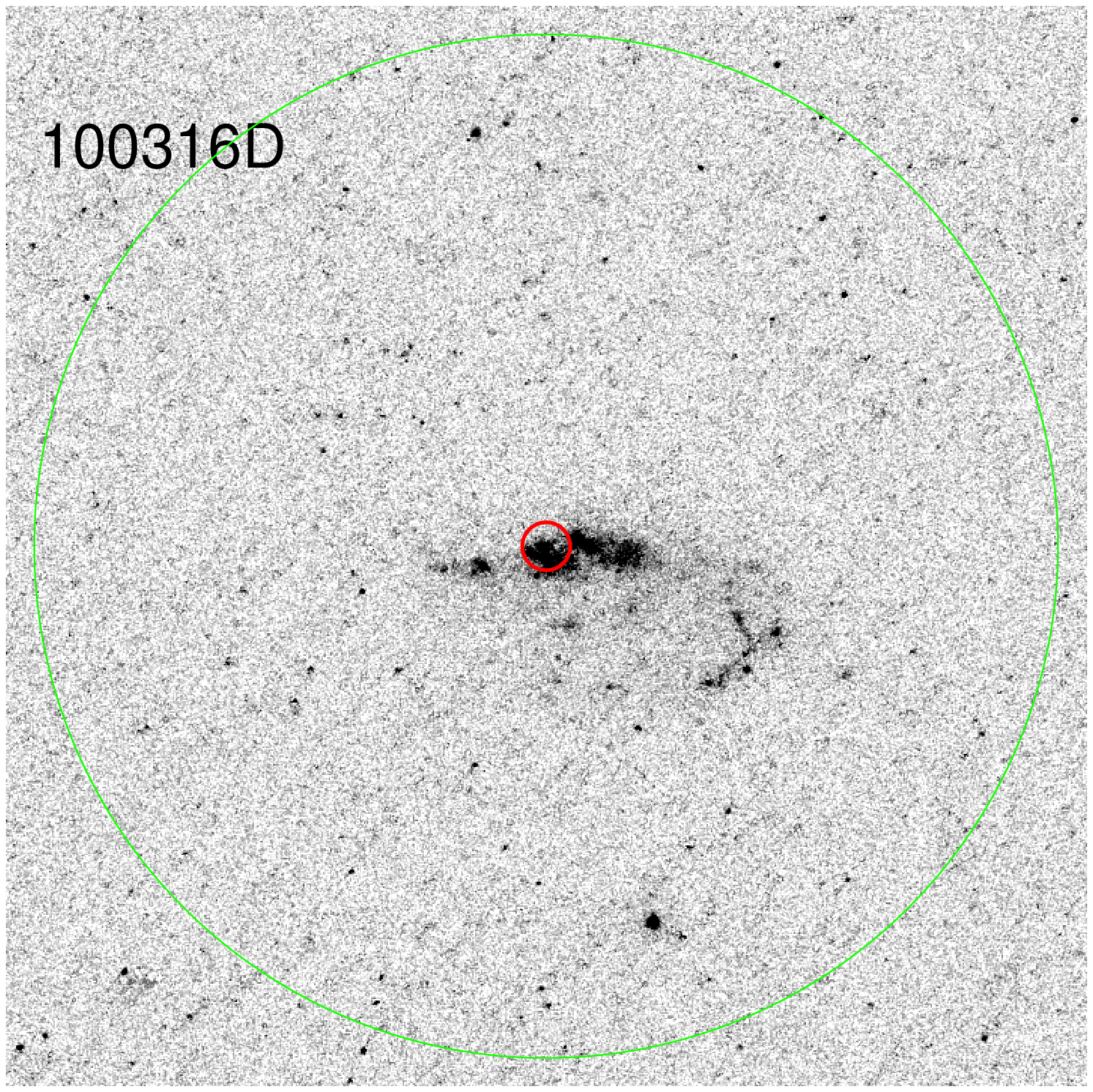} &
        \includegraphics[width= \mycolwi]{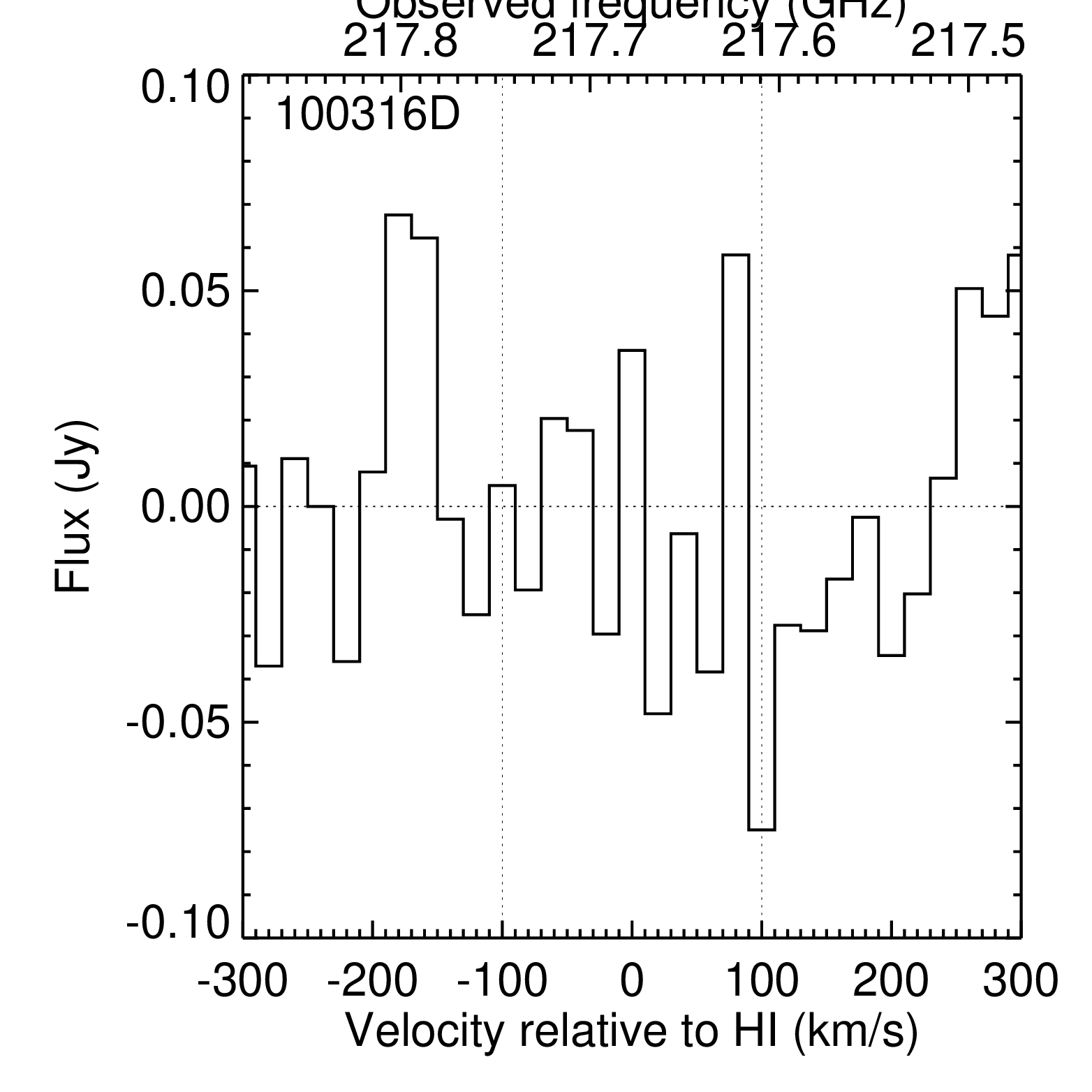}\\
\includegraphics[width= \mycolwi]{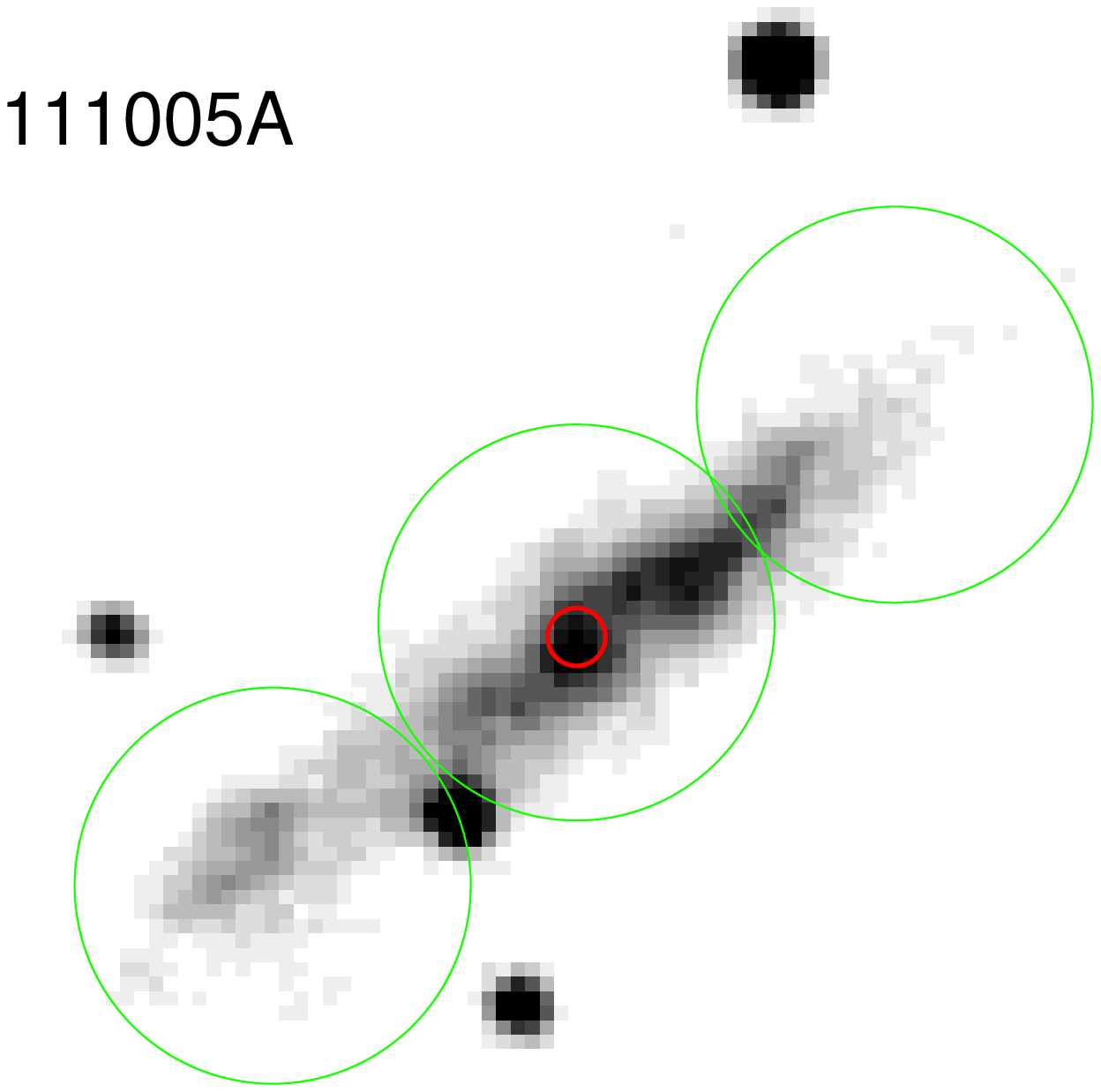} &
        \includegraphics[width= \mycolwi]{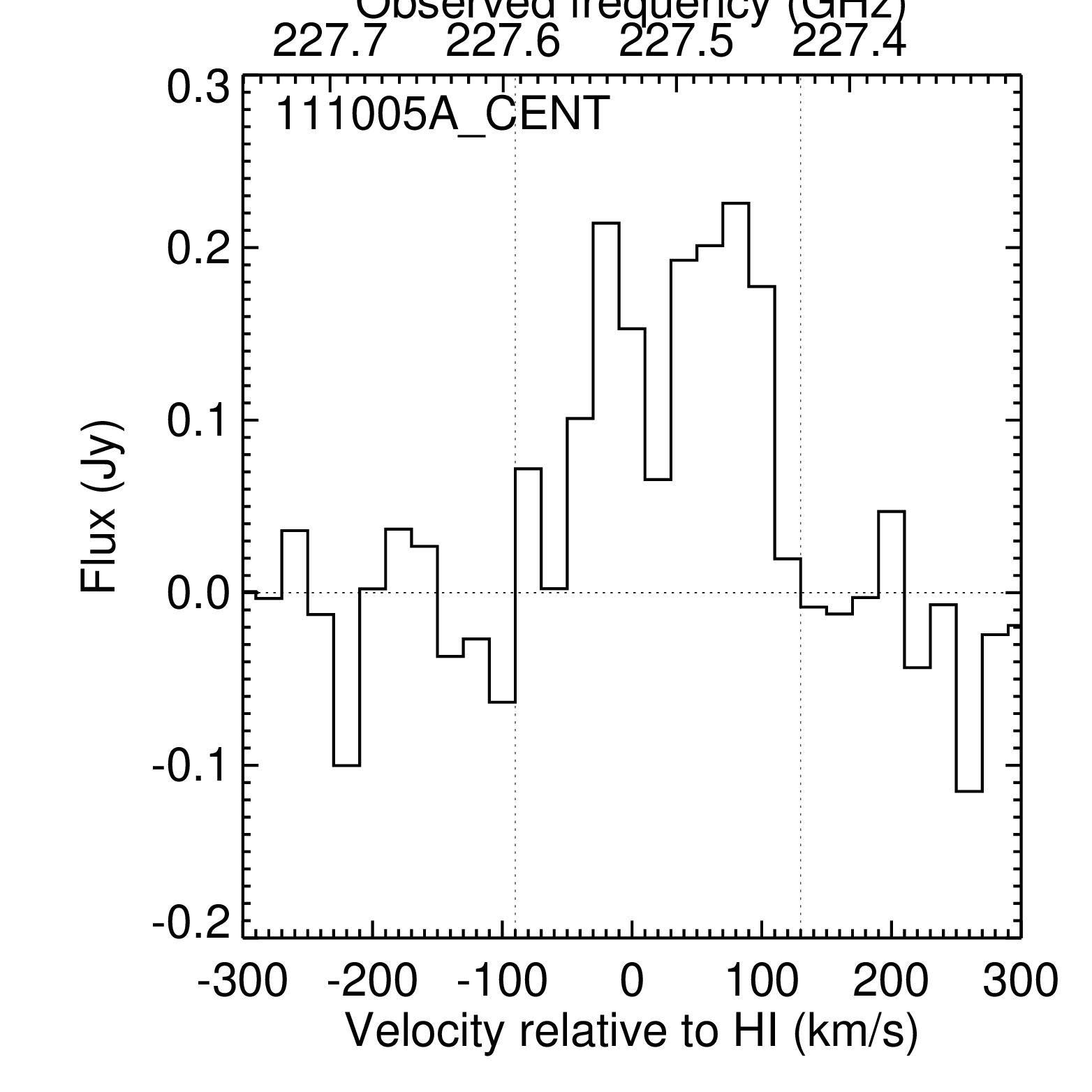} &
        \includegraphics[width= \mycolwi]{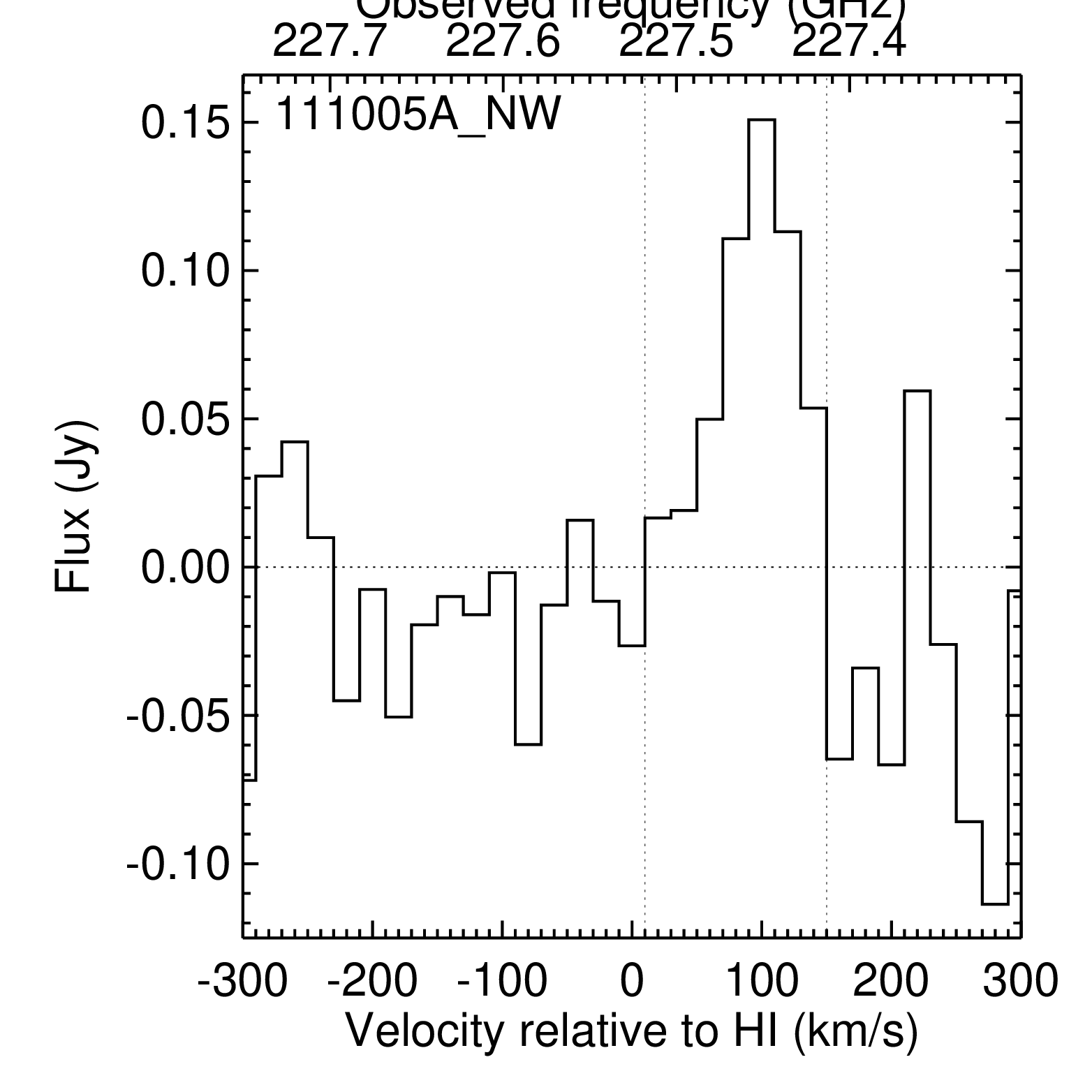} &
        \includegraphics[width= \mycolwi]{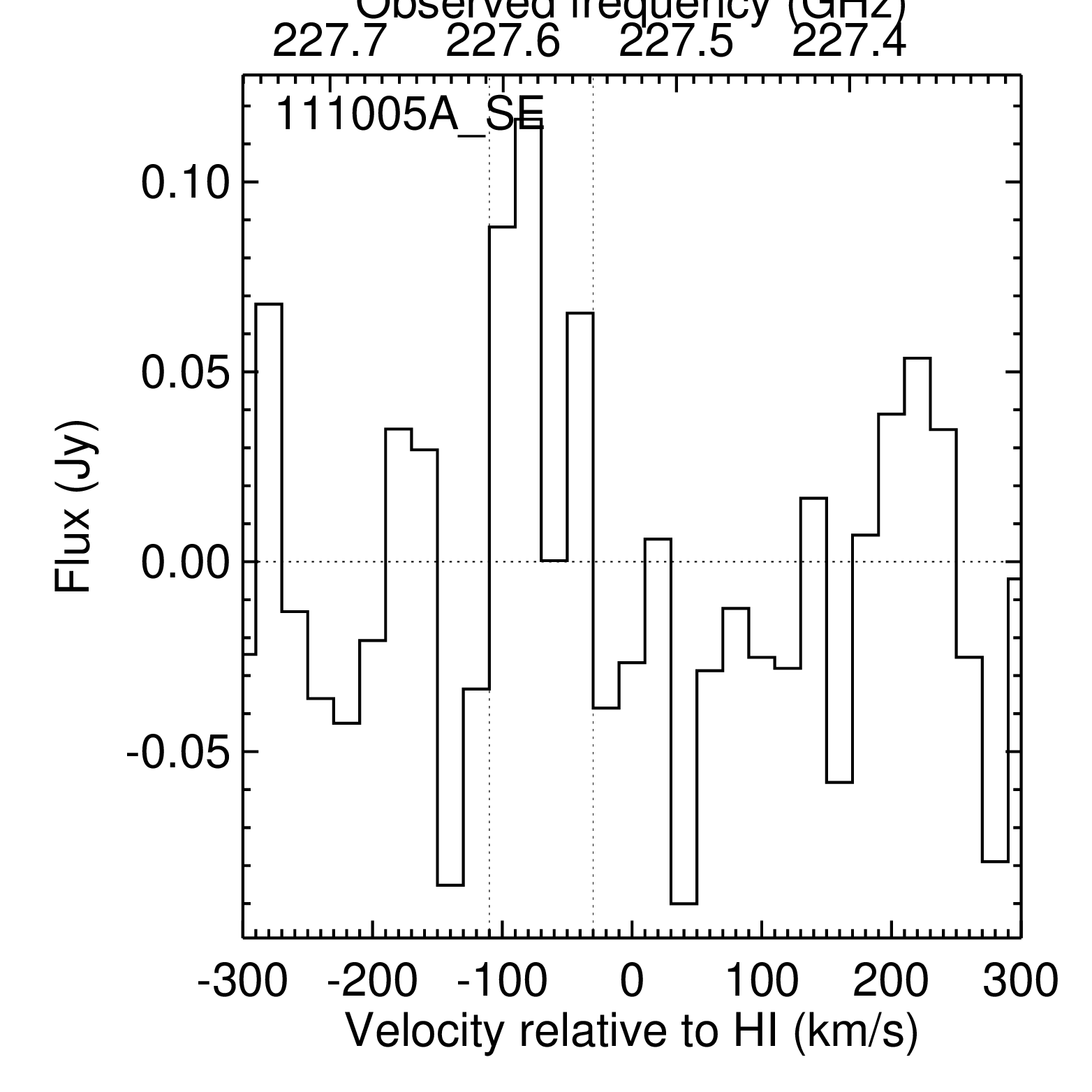}\\
\end{tabular}
\end{center}
\caption{
For each GRB host (labelled in the top left corner of each panel), the first panel shows the optical image \citep{%
sollerman05, 
mazzali06, 
thone08, 
hjorth12, 
starling11, 
michalowski18grb
} together with the {\it green circles} marking the positions of the pointings and the beam sizes of our CO(2-1) observations. GRB positions are marked with {\it red circles}. North is up and east is to the left. The other panels show the corresponding CO(2-1) spectra. {\it Vertical dotted lines} show the velocity intervals within which the line fluxes were measured.
}
\label{fig:spec}
\end{figure*}

\subsection{APEX}

We selected the host galaxies of all known
GRBs at $z < 0.12$ in the southern hemisphere \citep[i.e.~the sample with {\hi} observations from][]{michalowski15hi}. These criteria were fulfilled by
GRB 980425 \citep[the central pointing was published separately in][]{michalowski16}, 031203, 060505, 100316D, and 111005A. 
We performed CO(2-1) observations using the Swedish Heterodyne Facility Instrument \citep[SHeFI;][]{shefi,shefi2} and the Swedish-ESO PI Instrument for APEX (SEPIA; \citealt{sepia}; only for the GRB\,031203 host) mounted at the Atacama Pathfinder Experiment \citep[APEX;][]{apex}  (project no.~096.D-0280, 096.F-9302 and 097.F-9308, PI: M.~Micha{\l}owski). 
Table~\ref{tab:apexobs} shows the observation log with total on-source integration times.
Two and three positions were observed for the host of GRB\,980425 and 111005A, respectively. The remaining galaxies are smaller than the beam ($\sim27\arcsec$).
 All observations were carried out in the  on-off pattern and position-switching mode. The fluxes were corrected using the main beam efficiency of 0.75.
We reduced and analysed the data using the Continuum and Line Analysis Single Dish Software ({\sc Class}) package within the Grenoble Image and Line Data Analysis Software\footnote{\url{http://www.iram.fr/IRAMFR/GILDAS}} ({\sc Gildas}; \citealt{gildas}).

\begin{table}
\caption{Log of APEX observations.} \label{tab:apexobs}
\begin{tabular}{llcc}
\hline\hline
GRB & Obs.~Date & time/hr & pwv/mm \\
\hline
980425 Center & Total & 4.04 \\
                        & 2015 Aug 29 & 0.70 & 1.64--1.70 \\
                        & 2015 Sep 12 & 0.30 & 0.75--0.85 \\
                        & 2015 Sep 16 & 0.70 & 1.43--1.57 \\
                        & 2015 Oct 31 & 1.17 & 1.22--1.96 \\
                        & 2015 Nov 01 & 1.17 & 0.66--0.85 \\
\hline
980425 WR & Total & 6.57 \\
                   & 2015 Nov 02 & 2.17  & 0.75--3.48 \\
                   & 2016 Apr 03 & 0.10  & 2.02--2.15 \\
                   & 2016 Apr 04 & 4.30  & 3.33--5.23 \\
\hline
031203 & 2015 Sep 10 & 0.80  & 0.83--0.91 \\
\hline
060505 & Total & 7.00 \\
            & 2015 Aug 28 & 1.20 & 1.50-1.67 \\
            & 2015 Aug 29 & 1.40 & 1.38--1.62 \\
            & 2015 Sep 02 & 1.40  & 1.55--1.86 \\
            & 2015 Sep 03 & 1.00  & 3.36--3.61 \\
            & 2015 Sep 04 & 1.00  & 2.50--2.73 \\
            & 2015 Sep 06 & 1.00  & 2.45--3.40 \\
\hline
100316D & Total & 6.58 \\
               & 2015 Aug 28 & 2.11 & 1.50--1.62 \\
               & 2015 Sep 02 & 1.67  & 1.32--1.93 \\
               & 2015 Sep 06 & 2.80  & 2.45--4.80 \\
\hline
111005A Center & Total & 1.65 \\
                         & 2015 Sep 01 & 0.75  & 1.00--1.21 \\
                         & 2015 Sep 12 & 0.20  & 0.72--0.84 \\
                         & 2015 Sep 15 & 0.70  & 0.64--0.82 \\
\hline
111005A NW & Total & 3.20 \\
                     & 2015 Sep 17 & 0.50  & 1.52--1.61 \\
                     & 2016 Apr 02 & 1.00  & 2.15--2.47 \\
                     & 2016 Apr 03 & 0.60  & 1.96--2.31 \\
                     & 2016 Jun 10 & 1.60  & 2.98--3.34 \\
\hline
111005A SE & Total & 2.20 \\
                   & 2015 Sep 17 & 0.50  & 1.55--1.65 \\
                   & 2016 Jun 10 & 0.60 & 3.12--3.32 \\
                   & 2016 Jun 11 & 1.60 & 2.49--2.83 \\
\hline
\end{tabular}
\end{table}

\subsection{IRAM30m}

\begin{table}
\caption{Log of IRAM 30m observations.} \label{tab:iramobs}
\begin{tabular}{llcc}
\hline\hline
GRB & Obs.~Date & time/hr & $\tau_{225\,{\rm GHz}}$ \\
\hline
060814 & Total & 13.10  \\
             & 2017 Feb 01 & 0.40 & 0.29 \\
             & 2017 Feb 03 & 1.60 & 0.08--0.23 \\
             & 2017 Feb 04 & 3.20 & 0.23--0.51 \\
             & 2017 Feb 07 & 1.40 & 0.20--0.39 \\
             & 2017 Apr 06 & 0.70 & 0.13--0.17 \\
             & 2017 Apr 07 & 2.00 & 0.12--0.20 \\
             & 2017 Apr 08 & 2.20 & 0.15--0.19 \\
             & 2017 Apr 09 & 1.60 & 0.10--1.60 \\
\hline
080207 & Total & 17.80  \\
             & 2017 Feb 01 & 1.60 & 0.28--0.37 \\
             & 2017 Apr 11 & 1.90 & 0.27--0.36 \\
             & 2017 Apr 12 & 3.70 & 0.23--0.48 \\
             & 2017 Apr 13 & 4.30 & 0.20--0.44 \\
             & 2017 Apr 14 & 3.30 & 0.22--0.41 \\
             & 2017 May 22 & 3.00 & 0.24--0.36 \\
\hline
\end{tabular}
\end{table}

We selected all GRB hosts in the northern hemisphere with infrared or radio detections \citep{hunt14,perley15,michalowski15hi} and $z>1.5$, so that the line is located at lower frequencies and easier to observe. This was fulfilled by GRB\,060814 and 080207.
We performed observations with the IRAM 30m telescope (project no.~172-16, PI: M.~Micha{\l}owski) using the Eight MIxer Receiver\footnote{\url{www.iram.es/IRAMES/mainWiki/EmirforAstronomers}} \citep[EMIR;][]{emir}. 
We implemented wobbler-switching mode (with the offset to the reference positions of 60\arcsec), which provides stable and flat baselines and optimises the total observing time.
An intermediate frequency (IF) covered the frequency of the CO(2-1) line.
We used the Fourier Transform Spectrometers 200 (FTS-200) providing 195\,kHz spectral resolution (corresponding to $\sim0.8\,\kms$ at the frequency of CO(2-1) of our targets) and 16\,GHz bandwidth in each linear polarisation. 
The observations were divided into 6 min scans, each consisting of 12 scans 30\,s long. The pointing was verified every 1--2 hr. 
The  observing log is presented in Table~\ref{tab:iramobs} with total on-source integration times. The observations were carried out during good atmospheric conditions, and the opacity ($\tau_{\rm 225 GHz}$) was uniform across different runs.
We reduced the data using the {\sc Class} package within {\sc Gildas} \citep{gildas}.
Each spectrum was calibrated, and corrected for baseline shape. 
The spectra were aligned in frequency and noise-weight averaged. Some well-known platforming, due to the fact that the instantaneous bandwidth of 4 GHz is sampled by three different FTS units, was corrected off-line by a dedicated procedure within {\sc Class}. In all cases, the CO line is far away from the step of the platforming.


\begin{figure*}
\begin{center}
\includegraphics[width=\textwidth]{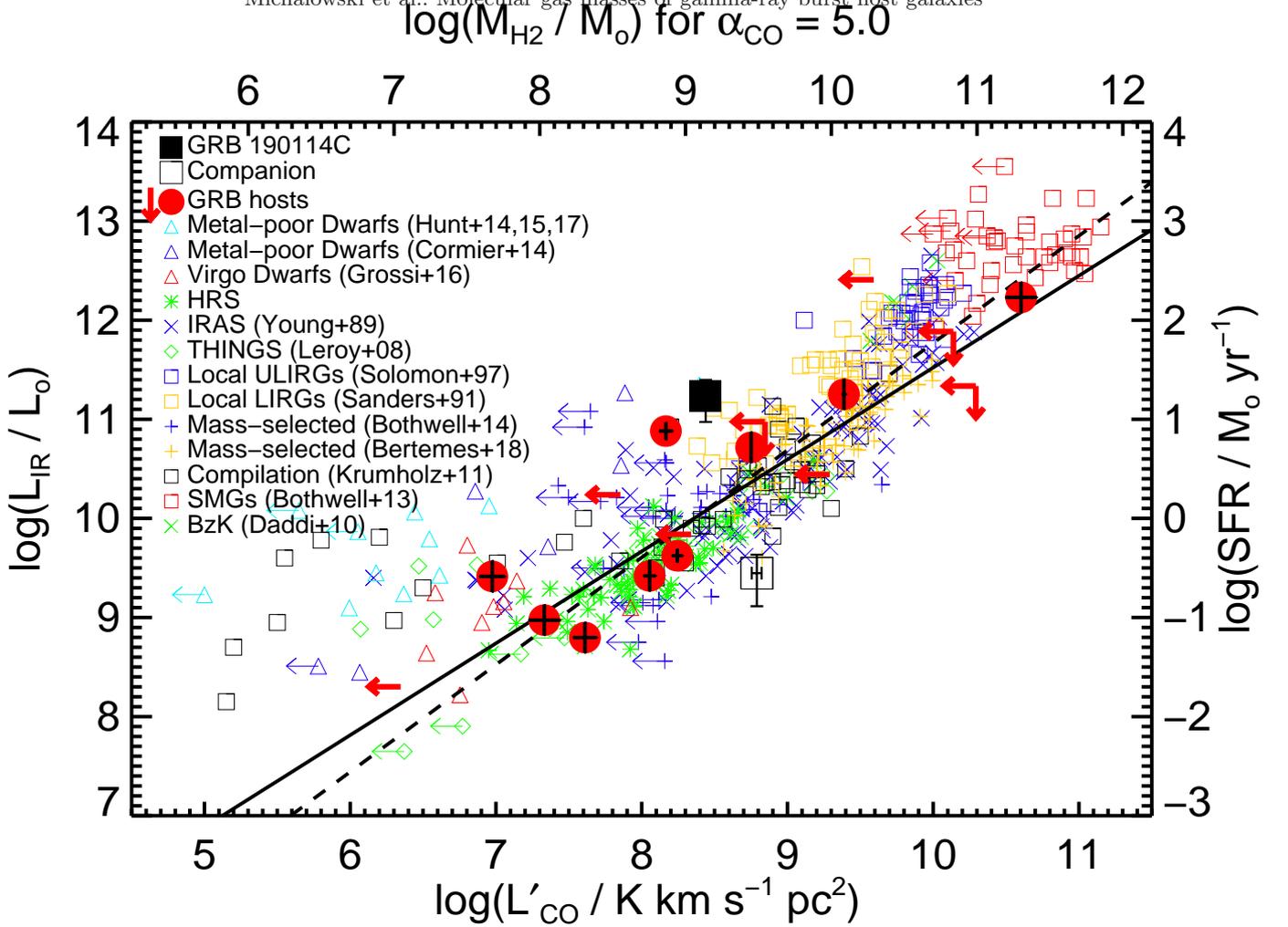}
\end{center}
\caption{Infrared luminosity or the corresponding SFR as a function of CO luminosity, or the corresponding molecular gas mass with the CO-to-{\htwo} conversion factor $\alpha_{\rm CO}=5\,M_\odot\, (\mbox{K km s}^{-1} \mbox{ pc}^2)^{-1}$. GRB hosts are marked with {\it full red circles or red arrows} with crosses showing the errors. 
The symbols of other galaxies are indicated in the legend and described in Sect.~\ref{sec:other}. 
The {\it solid black line} is a linear fit to the non-GRB galaxies excluding ULIRGs (Eq.~\ref{eq:mh2sfr}), 
whereas  the {\it dashed black line} represents the fit including ULIRGs (Eq.~\ref{eq:mh2sfrulirg}).
The $\sim0.3$\,dex shift for GRB hosts towards lower {\mhtwo} is not statistically significant (see Sect.~\ref{sec:sfrmh2}).
}
\label{fig:sfrmh2}
\end{figure*}

\subsection{Literature data for additional GRB hosts}
\label{sec:othergrb}

In addition to the CO(2-1) measurements obtained here, we included all other CO measurements for GRB hosts from the literature. All molecular masses were converted into $\alpha_{\rm CO}=5\,M_\odot\, (\mbox{K km s}^{-1} \mbox{ pc}^2)^{-1}$ and to the line luminosity ratios in temperature units $\lpcotwo/\lpcoone=0.5$, $\lpcothree/\lpcoone=0.27$, or $\lpcofour/\lpcoone=0.17$ \citep[the Milky Way values, see Table 2 of][]{carilli13} if these masses were based on CO(2-1), CO(3-2), or CO(4-3) observations, respectively.
These assumed line ratios are conservatively low, so that they lead to conservatively high {\mhtwo}. We are therefore able to robustly test for any molecular deficiency of GRB hosts.

We included the hosts of 
GRB\,000418 \citep[][]{hatsukade11b}, for which we converted the $\mhtwo$ upper limit from  $\lpcotwo/\lpcoone=1$ into $0.5$ and from $\alpha_{\rm CO}=0.8\,M_\odot\, (\mbox{K km s}^{-1} \mbox{ pc}^2)^{-1}$ to $5$; of GRB
030329 \citep{kohno05,endo07}, for which we converted the $\mhtwo$ upper limit from  $\alpha_{\rm CO}=40\,M_\odot\, (\mbox{K km s}^{-1} \mbox{ pc}^2)^{-1}$ into $5$; of GRB
051022 \citep[][]{hatsukade14}, for which we converted the $\mhtwo$ detection from  $\lpcofour/\lpcoone=0.85$ into $0.17$ and from $\alpha_{\rm CO}=4.3\,M_\odot\, (\mbox{K km s}^{-1} \mbox{ pc}^2)^{-1}$ into $5$; of GRB
080517 \citep[][]{stanway15}, for which we converted the $\mhtwo$ detection from $\alpha_{\rm CO}=4.3\,M_\odot\, (\mbox{K km s}^{-1} \mbox{ pc}^2)^{-1}$ into $5$;
and of GRB 090423 \citep[][]{stanway11}, for which we converted the $\mhtwo$ detection from  $\lpcothree/\lpcoone=1$ into $0.27$ and from $\alpha_{\rm CO}=0.8\,M_\odot\, (\mbox{K km s}^{-1} \mbox{ pc}^2)^{-1}$ into $5$.

We did not use the CO(3-2) observations of GRB\,980425 of \citet{hatsukade07} because our deeper data resulted in a detection. Moreover, we excluded GRB\,020819B 
because the low-redshift galaxy with the existing CO measurement \citep{hatsukade14} has been shown not to be related to the GRB \citep{perley17}.
For the GRB\,080207 host, the CO(3-2) line observations were recently reported by \citet{arabsalmani18}. We did not use these values in subsequent analysis, because our lower transition likely traces a larger fraction of the total molecular gas content.
We note, however, that the obtained gas masses are consistent (see Sect.~\ref{sec:res}).

For all GRB hosts in our CO sample we used the literature values for their redshifts, 
SFRs 
and metallicities, 
as listed in Table~\ref{tab:sample}
For the host of GRB\,060814, we calculated the metallicity based on the $R_{23}$ method of \citet{kobulnicky04} based on the [\ion{O}{ii}], [\ion{O}{iii}], and H$\beta$ emission lines, using the fluxes reported in \citet{kruhler15}. We obtained $\metoh\sim8.38\pm0.35$.

Additionally, we included values measured for the host of SN\,2009bb, the relativistic supernova (SN) type Ic \citep{michalowski18} and plot them in Figs.~\ref{fig:sfrmh2} and \ref{fig:sfrmh2metal}. SNe of this type may have similar engines as GRBs, but no $\gamma$-rays were detected. Therefore we did not use it for the statistical analysis quoted for GRB hosts, and it does not appear in Figs.~\ref{fig:sfrmh2cumul} and \ref{fig:sfrmh2metalcumul}.

\subsection{Other galaxy samples}
\label{sec:other}

In order to place the GRB hosts in the context of general galaxy populations, we compared their properties with those of the following galaxy samples, chosen based on the availability of the gas mass estimates:
the optical-flux-limited spirals and irregulars with IRAS data \citep{young89b},
local luminous infrared galaxies \citep[LIRGs;][]{sanders91},
local ultra-luminous infrared galaxies   \citep[ULIRGs;][]{solomon97},
the {\it Herschel} Reference Survey \citep[HRS;][]{boselli10,cortese12b,cortese14,boselli14,ciesla14},
 {\hi}-dominated, low-mass galaxies and large spiral galaxies \citep{leroy08},
$0.01<z<0.03$ mass-selected galaxies with $8.5<\log(\mstar/\msun)<10$ \citep{bothwell14},
$0.025 < z < 0.2$ mass-selected galaxies with $\log(\mstar/\msun)>10$ and infrared detections \citep{bertemes18},
%
metal-poor dwarfs \citep{hunt14b,hunt15,hunt17,leroy07},
metal-poor dwarfs from the {\it Herschel} Dwarf Galaxy Survey \citep{madden13,cormier14},
Virgo-cluster dwarfs \citep{grossi16},
$z\sim1.5$ BzK galaxies \citep{daddi10,magdis11,magnelli12b},
and $1.2<z<4.1$ {\smgs} \citep{bothwell13,michalowski10smg}.

All 
SFRs were converted into the \citet{chabrier03} IMF. The molecular masses were converted into $\alpha_{\rm CO}=5\,M_\odot\, (\mbox{K km s}^{-1} \mbox{ pc}^2)^{-1}$ and to the Milky Way line ratios if they were based on higher CO transitions. Namely, \citet{bothwell14}, \citet{daddi10}, and \citet{leroy08} assumed $\lpcotwo/\lpcoone=1$, $0.16$, and $0.8$ respectively, and \citet{hunt14b} assumed $\lpcothree/\lpcoone=0.6$.
 The Galactic value of $\alpha_{CO}$ is appropriate for $0.4$--$1$ solar metallicity galaxies discussed here \citep{bolatto13,hunt14b}.
Following \citet{hunt15}, metallicities from \citet{bothwell14} were converted from the calibration of \citet[][KD02]{kewley02} into that of \citet[][PP04 N2]{pettini04} using the equation derived by \citet[][their Table 3]{kewley08}.

Even though SFR estimates of other galaxies are often derived from various diagnostics (ultraviolet, H$\alpha$, infrared, and radio), they are broadly consistent \citep{salim07,wijesinghe11,davies16,wang16}, even in dwarf galaxies, except at very low $\mbox{SFR}<0.001\msunyr$ \citep{huang12,lee09b}, not discussed here.

\section{Results}
\label{sec:res}

\begin{table*}
\centering
\caption{APEX and IRAM 30m CO(2-1) line fluxes and luminosities.\label{tab:co}}
\medskip
\begin{tabular}{lrrrrrr}
\hline\hline
GRB & \multicolumn{1}{c}{$F_{\rm int}$} & S/N & \multicolumn{1}{c}{$F_{\rm int}$} & \multicolumn{1}{c}{$\log L$} & \multicolumn{1}{c}{$\log \lp$} & \multicolumn{1}{c}{$\log M_{\rm H_2, CO}$}   \\
       & \multicolumn{1}{c}{(Jy km s$^{-1}$)} &  & \multicolumn{1}{c}{($10^{-20}$W m$^{-2}$)}    & \multicolumn{1}{c}{($L_\odot$)} & \multicolumn{1}{c}{($\mbox{K km s}^{-1} \mbox{ pc}^2$)} &\multicolumn{1}{c}{($\msun$)}   \\
(1)    & \multicolumn{1}{c}{(2)}     & (3)  & \multicolumn{1}{c}{(4)}           & \multicolumn{1}{c}{(5)} & \multicolumn{1}{c}{(6)} & \multicolumn{1}{c}{(7)} \\
\hline
980425 & $ 5.65\pm1.27$ & $ 4.4$ & $ 4.34\pm0.98$ & $ 3.27_{-0.11}^{+0.09}$ & $ 6.67_{-0.11}^{+0.09}$ & $ 7.67_{-0.11}^{+0.09}$ \\
980425\_WR & $ 1.33\pm1.28$ & $ 1.0$ & $ 1.02\pm0.98$ & $ 2.64_{-1.43}^{+0.29}$ & $ 6.04_{-1.43}^{+0.29}$ & $ 7.04_{-1.43}^{+0.29}$ \\
031203 & $ 7.51\pm3.35$ & $ 2.2$ & $ 5.77\pm2.58$ & $ 5.58_{-0.26}^{+0.16}$ & $ 8.99_{-0.26}^{+0.16}$ & $ 9.99_{-0.26}^{+0.16}$ \\
060505 & $ 1.18\pm1.64$ & $ 0.7$ & $ 0.91\pm1.26$ & $<$$ 5.21$ & $<$$ 8.62$ & $<$$ 9.62$ \\
060814 & $ -0.04\pm0.11$ & $ -0.4$ & $ -0.03\pm0.09$ & $<$$ 6.51$ & $<$$ 9.92$ & $<$$ 10.92$ \\
080207 & $ 0.38\pm0.11$ & $ 3.5$ & $ 0.29\pm0.08$ & $ 6.90_{-0.14}^{+0.11}$ & $ 10.30_{-0.14}^{+0.11}$ & $ 11.30_{-0.14}^{+0.11}$ \\
100316D & $ -0.88\pm2.25$ & $ -0.4$ & $ -0.68\pm1.73$ & $<$$ 4.76$ & $<$$ 8.16$ & $<$$ 9.16$ \\
111005A\_CENT & $ 28.49\pm2.94$ & $ 9.7$ & $ 21.91\pm2.26$ & $ 4.35_{-0.05}^{+0.04}$ & $ 7.75_{-0.05}^{+0.04}$ & $ 8.75_{-0.05}^{+0.04}$ \\
111005A\_NW & $ 10.27\pm2.13$ & $ 4.8$ & $ 7.90\pm1.63$ & $ 3.90_{-0.10}^{+0.08}$ & $ 7.31_{-0.10}^{+0.08}$ & $ 8.31_{-0.10}^{+0.08}$ \\
111005A\_SE & $ 5.41\pm1.52$ & $ 3.5$ & $ 4.16\pm1.17$ & $ 3.62_{-0.14}^{+0.11}$ & $ 7.03_{-0.14}^{+0.11}$ & $ 8.03_{-0.14}^{+0.11}$ \\
\hline
\end{tabular}
\tablefoot{
(1) GRB (2) Integrated flux within the velocity interval shown by the dotted lines in Fig.~\ref{fig:spec}. (3) Signal-to-noise ratio of the line within this velocity interval. (4) Corresponding integrated flux in W m$^{-2}$. (5) Line luminosity. (6) Line luminosity in temperature units based on Equation 3 in \citet{solomon97}. (7) Molecular gas mass estimated assuming $L'_{\rm CO(1-0)}= 2\times L'_{\rm CO(2-1)}$ (see Sects.~\ref{sec:othergrb} and \ref{sec:res}) and the Galactic CO-to-{\htwo} conversion factor $\alpha_{\rm CO}=5\msun/\Kkmspc$.
}
\end{table*}

\begin{figure}
\begin{center}
\includegraphics[width=0.5\textwidth]{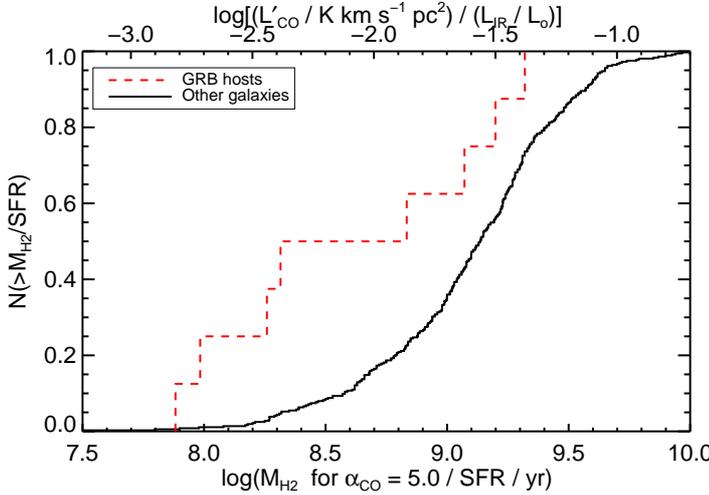}
\end{center}
\caption{Cumulative distribution of molecular gas depletion time (or the inverse of the star formation efficiency), i.e.~the ratio of the CO luminosity to the infrared luminosity or the corresponding molecular gas mass with the CO-to-{\htwo} conversion factor $\alpha_{\rm CO}=5\,M_\odot\, (\mbox{K km s}^{-1} \mbox{ pc}^2)^{-1}$ to the star formation rate (SFR). The distribution of GRB hosts is shown as the {\it dashed red line}, 
whereas that of other galaxies is shown as the {\it solid black line}.
We treated the upper limits as actual values, so the histogram for GRB hosts is an upper limit.
GRB hosts are systematically shifted to the left on this diagram (lower {\mhtwo} given their SFRs),
but this is not statistically significant (see Sect.~\ref{sec:sfrmh2}).
}
\label{fig:sfrmh2cumul}
\end{figure}

The positions of our APEX and IRAM 30m pointings and the obtained CO(2-1) spectra are shown in Fig.~\ref{fig:spec}. 
The spectra were binned to a velocity resolution of 20\,{\kms}, except for the GRB\,080207 host, for which 50\,{\kms} channels were adopted.
 The derived parameters are shown in Table~\ref{tab:co}. The fluxes were integrated within the velocity ranges shown  in Fig.~\ref{fig:spec} as vertical dotted lines.
They were chosen to encompass the full extent of the lines for the detected targets, and the most significant positive feature within the velocity range from $-300$ to $300\,\kms$ relative to the optical redshift for the non-detected targets in order to obtain the most conservative upper limits.
For these non-detected targets we integrated the spectra in the region of a width of $200\,\kms$, likely to be the velocity width of such galaxies, and of $50\,\kms$ for the WR region, as it is unlikely that this pointing traces gas at a wider range of velocities \citep[see Fig.~3 of][]{christensen08}.
 The CO(2-1) line luminosities were calculated using Equation 3 in \citet{solomon97} and converted into the CO(1-0) luminosities assuming $\lpcoone= 2\times \lpcotwo$. The Galactic CO-to-{\htwo} conversion factor $\alpha_{\rm CO}=5\msun/\Kkmspc$ was used to calculate molecular gas masses ($\mhtwo=\alpha_{\rm CO}\lpcoone$).


\subsection{SFR vs.~{\mhtwo}}
\label{sec:sfrmh2}

The infrared luminosity (or SFR) as a function of CO line luminosity (or {\mhtwo}) for GRB hosts and other galaxies is shown in Fig.~\ref{fig:sfrmh2}. 
The best linear fit in log-log space to all non-GRB galaxies with SFRs lower than those of ULIRGs ($\mbox{SFR}<100\,\msunyr$) is (the solid line in Fig.~\ref{fig:sfrmh2})
\begin{equation}
\log(\mbox{SFR}/\msunyr)=0.95\times\log({\mhtwo}/\msun) -8.57
\label{eq:mh2sfr}
.\end{equation}
The scatter around this relation is $\sim0.42$\,dex.
When ULIRGs are included, this equation changes to (the dashed line in Fig.~\ref{fig:sfrmh2})
\begin{equation}
\log(\mbox{SFR}/\msunyr)=1.10\times\log({\mhtwo}/\msun) -9.96
\label{eq:mh2sfrulirg}
.\end{equation}

As reported in \citet{michalowski16}, we found a low molecular gas content in the GRB\,980425 host given its SFR. Similarly, the hosts of GRB\,100316D and 060814 are deficient in {\mhtwo} given their SFRs.
Our {\mhtwo} upper limit for the GRB\,031203 host is 
$\sim0.5$\,dex higher than the value suggested by the best-fit relation of Eq.~(\ref{eq:mh2sfr})
so that we cannot conclude much about its molecular gas content. 
 \citet{wiersema18} measured a molecular gas mass $\sim5$ times lower than our upper limit based on the $H_2$ 0-0 S(7) rotational emission line.
Our {\mhtwo} upper limit for the GRB\,060505 host is not 
sufficiently strong to test for any
molecular gas deficiency, but it is close to the best-fit line for other star-forming galaxies, which means that this galaxy is not richer in molecular gas than the average of other galaxies.
We found that the GRB\,080207 host is very close to the best-fit line for other galaxies on the SFR-{\mhtwo} diagram, consistent with the results of \citet{arabsalmani18} based on the CO(3-2) line.
The host of GRB\,111005A is molecule rich with $\log({\mhtwo}/\mbox{SFR}/\mbox{yr})\sim9.34$, that is, $\sim0.24$\,dex above the best-fit relation for other galaxies at the relevant SFR. Consistently with \citet{michalowski18}, we show that the host of SN\,2009bb has a molecular gas mass that is a few times lower than its SFR suggests.

The second pointing for the GRB\,980425 host, towards the Wolf-Rayet (WR) region \citep[for its properties, see][]{hammer06,lefloch,lefloch12,christensen08,michalowski09,michalowski14,michalowski16,kruhler17} resulted in an upper limit close to the best-fit line. 
While we cannot establish any molecular deficiency for this region, it is therefore definitely not molecule rich, in contrast with its high abundance of atomic gas \citep{arabsalmani15b}.
Both the central and NW regions of the GRB\,111005A host are molecule rich, but the SE region is at least slightly molecule deficient, 
given its CO upper limit.

\begin{figure*}
\begin{center}
\includegraphics[width=\textwidth]{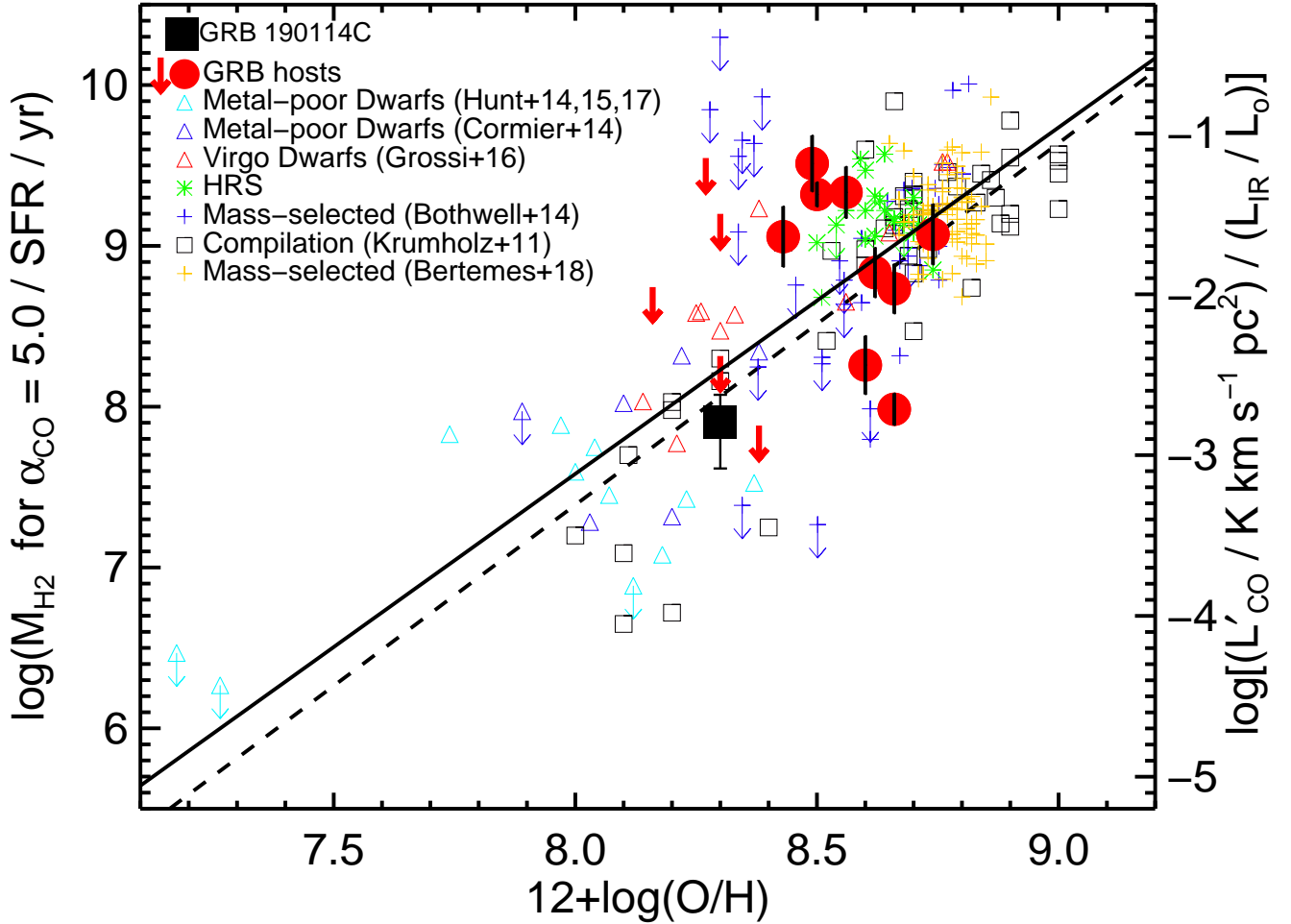}
\end{center}
\caption{Molecular gas depletion time (or the inverse of the star formation efficiency), i.e.~the ratio of the CO luminosity to the infrared luminosity or the corresponding molecular gas mass with the CO-to-{\htwo} conversion factor $\alpha_{\rm CO}=5\,M_\odot\, (\mbox{K km s}^{-1} \mbox{ pc}^2)^{-1}$ to the SFR as a function of metallicity.  GRB hosts are marked with {\it full red circles or red arrows} with vertical bars showing the errors.
The symbols of other galaxies are indicated in the legend and described in Sect.~\ref{sec:other}.
The {\it solid black line} is our fit to the non-GRB galaxies (eq.~\ref{eq:mh2sfrmetal}), whereas the {\it dashed black line} is the relation found by \citet{hunt15}.
GRB hosts are consistent with other galaxies (see Sect.~\ref{sec:sfrmh2metal}).
}
\label{fig:sfrmh2metal}
\end{figure*}

Because of our choice to adopt the Milky Way CO line ratios instead of those of M82 (see Sect.~\ref{sec:othergrb}), we obtained a molecular gas mass that is approximately five times higher for the GRB\,051022 host, and hence its molecular gas deficiency is not as dramatic as presented originally in \citet{hatsukade14}, but still apparent (Fig.~\ref{fig:sfrmh2}). 
Our correction for the GRB\,080517 is small with respect to the values used in \citet{stanway15}, so we recover its reported molecular gas deficiency. 

The revised, lower value of the infrared luminosity of the host of GRB\,000418 (compare \citealt{michalowski08} and \citealt{perley17}) means that the CO observations \citep{hatsukade11b} do not provide useful constraints on its location on the SFR-{\mhtwo} diagram (see Fig.~\ref{fig:sfrmh2}). 
Similarly, the upper limits on $L_{\rm IR}$ available for GRB\,030329 \citep{endo07} and 090423 \citep{stanway11} do not constrain the positions of these galaxies relative to the best-fit SFR-{\mhtwo} relation. Hence we did not use these three hosts with upper limits for both SFRs and {\mhtwo} in the statistical analysis.

The median value of the molecular gas depletion time for non-GRB galaxies is $\log(\mhtwo/\mbox{SFR}/\mbox{yr})=9.099^{+0.031}_{-0.020}$, whereas for GRB hosts it is  $8.83^{+0.24}_{-0.52}$ (the errors were obtained by randomly perturbing 500 times the measured values within their errors and assessing the 68\% confidence interval of the obtained medians),  
where we treated the upper limits as actual values. The value for GRB hosts therefore is an upper limit. Hence, GRB hosts have molecular gas masses $\sim0.3$\,dex below the expectations from their SFR, but this result has low significance.

The cumulative distributions of the {\mhtwo}/SFR ratio (molecular gas depletion time) is shown in Fig.~\ref{fig:sfrmh2cumul}. 
For these statistics we 
excluded hosts with weak upper limits (031203)  and those with upper limits for both {\mhtwo} and SFRs (000418, 030329, and 090423).
Using the Kolmogorov--Smirnov (K-S) test, we 
found that we can rule out the null hypothesis that the $\mhtwo/\mbox{SFR}$ values of the GRB hosts were drawn from the same distribution as those of other star-forming galaxies at a significance level $p=0.05$,
corresponding to a difference with a low statistical significance of $\sim1.9\sigma$. 
In order to assess the influence of the measurement errors on this result, we repeated the K-S test using the GRB values perturbed by their errors and found that the significance remains similar.


\subsection{{\mhtwo}/SFR vs.~metallicity}
\label{sec:sfrmh2metal}

The CO-to-{\htwo} conversion factor is metallicity dependent \citep[e.g.][]{bolatto13}, therefore we explored the {\mhtwo}/SFR ratio as a function of metallicity (Fig.~\ref{fig:sfrmh2metal}). 
Using the galaxies with metallicity measurement, the linear fit to all non-GRB galaxies  is (the solid line in Fig.~\ref{fig:sfrmh2metal})
\begin{equation}
\log({\mhtwo}/\mbox{SFR}/\mbox{yr})=2.33\times[\metoh] -11.1
\label{eq:mh2sfrmetal}
.\end{equation}
The scatter around this relation is $\sim0.35$\,dex.

The molecular deficiency of the GRB\,980425 is confirmed, even taking into account its sub-solar metallicity, that is to say,~it has a shorter molecular gas depletion time than expected for its SFR and metallicity. This is at odds with the discussion in \citet{arabsalmani18} that this galaxy has normal molecular gas properties. However, they compared {\mhtwo} with stellar mass, not SFR, as we do here, and also used the dwarf sample of \citet{grossi16} as a comparison, but these galaxies exhibit much lower metallicities than the GRB\,980425 host (see Fig.~\ref{fig:sfrmh2metal}). Similarly, the molecular gas deficiency of the hosts of GRB\,080517 and 060814 is confirmed after taking into account their metallicities.

The hosts of GRB\,051022, 080207, and 100316D 
have depletion times consistent with the expected values given their metallicities (the GRB\,100316D host represents an upper limit, therefore we do not know whether it is close to the best-fit relation).
Only the GRB\,111005A host is clearly molecule rich for its metallicity.  The limits for the hosts of GRB\,031203 and 060505 are not constraining because they are significantly above the best fit line.

Our upper limit for the WR region of the GRB\,980425 host is $\sim0.4$\,dex above the best-fit line in Fig.~\ref{fig:sfrmh2metal}, 
but the beam size of our observations is much larger than this region (Fig.~\ref{fig:spec}), which means that in reality our observations also probe the higher-metallicity regions.

Similarly to the results presented in Sect.~\ref{sec:sfrmh2}, the central and NW regions of the GRB\,111005A host are rich in molecular gas given their SFR and metallicity. On the other hand, the SE region has a much lower molecular gas content, close to the best-fit line.

For GRB hosts, the median value of the residual from this best fit is  $-0.21\pm0.07\,\mbox{yr}^{-1}$, where we treated the upper limits as actual values. This value is therefore an upper limit.

\begin{figure}
\begin{center}
\includegraphics[width=0.5\textwidth]{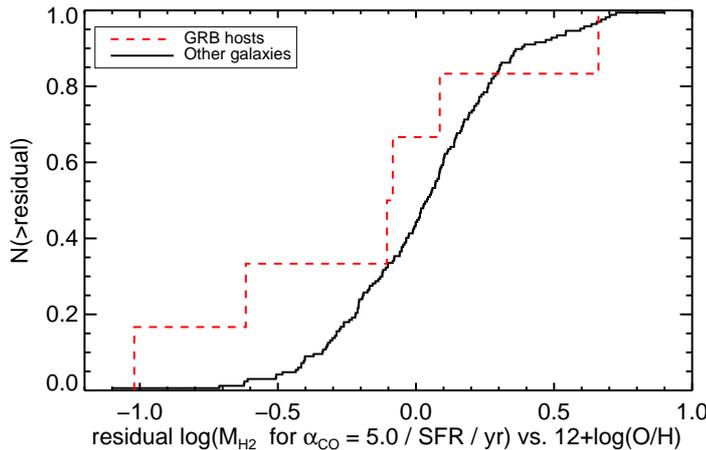}
\end{center}
\caption{Cumulative distribution of the residuals with respect to the solid line in Fig.~\ref{fig:sfrmh2metal} (Eq.~\ref {eq:mh2sfrmetal}), showing the relation between metallicity and molecular gas depletion time (or the inverse of the star formation efficiency), i.e.~the ratio of the CO luminosity to the infrared luminosity or the corresponding molecular gas mass with the CO-to-{\htwo} conversion factor $\alpha_{\rm CO}=5\,M_\odot\, (\mbox{K km s}^{-1} \mbox{ pc}^2)^{-1}$ to the SFR. The distribution of GRB hosts is shown as the {\it dashed red line}, 
whereas that of other galaxies is shown as the {\it solid black line}.
We treated the upper limits as actual values, so the histogram for GRB hosts is an upper limit.
GRB hosts are systematically shifted to the left on this diagram (lower {\mhtwo} given their SFRs and metallicity),
but this is not statistically significant (see Sect.~\ref{sec:sfrmh2metal}).
}
\label{fig:sfrmh2metalcumul}
\end{figure}

 The cumulative distributions of residuals around the best-fit line (Eq.~\ref{eq:mh2sfrmetal}) is shown in Fig.~\ref{fig:sfrmh2metalcumul}.
 For these statistics we 
 excluded hosts with weak upper limits (031203 and 060505)  and those with upper limits for both {\mhtwo} and SFRs (000418, 030329, and 090423).
Using the K-S test, we 
we found that we can reject the null hypothesis that the residuals around the best-fit line for GRB hosts were drawn from the same distribution as those for other star-forming galaxies only at a significance level $p=0.33$,
corresponding to a $\sim1\sigma$ difference. 


\subsection{Molecular gas fraction}

Using the {\hi} data from \citet{michalowski15hi}, we can constrain the molecular gas fraction ($\mhtwo/(\mhtwo + \mhi)$) to be $\sim7$\% for the GRB\,980425 host,
$<15$\% for the GRB\,060505 host,
 and $\sim13$\% for the GRB\,111005A host.
This is within the scatter of but on the lower side compared to other star-forming galaxies \citep[a few to a few tens of percent ;][]{young89b,devereux90,leroy08,saintonge11,cortese14,boselli14} and SN hosts \citep{galbany17, michalowski18}.

\section{Discussion}
\label{sec:discussion}

We obtained mixed results from analysing CO data for 12 GRB hosts from our survey and from the literature.
Three GRB hosts are clearly deficient in molecular gas, even taking into account their metallicity (980425, 060814, and 080517).
Four others are close to the best fit-line for other star-forming galaxies in the SFR-{\mhtwo} plot (051022, 060505, 080207, and 100316D). 
One host is clearly molecule-rich (111005A). 
Finally, for 4 GRB hosts the data are not deep enough to judge whether they are molecule deficient (000418, 030329, 031203, and 090423).

These results suggest that GRB hosts may be preferentially found in galaxies with lower molecular gas content than other star-forming galaxies, as there are more examples of GRB hosts in the {\mhtwo}-poor part of the {\mhtwo}-SFR diagram, and the median molecular depletion timescale ({\mhtwo}/SFR) of GRB hosts is $\sim0.3$\,dex shorter that of other galaxies. However, the difference between GRB hosts and other star-forming galaxies is significant only at the $\sim2\sigma$ level when analysing {\mhtwo}/SFR (Figs.~\ref{fig:sfrmh2} and \ref{fig:sfrmh2cumul}). Moreover, the statistical significance of this tentative difference decreases further to the $\sim1\sigma$ level when taking the metallicity into account  (Figs.~\ref{fig:sfrmh2metal} and \ref{fig:sfrmh2metalcumul}). Hence, our sample is statistically consistent with other star-forming galaxies.

Recent high-resolution observations of GRB and SN hosts showed concentrations of atomic gas close to the GRB and SN positions \citep{michalowski15hi,michalowski18,arabsalmani15b}, 
strongly supporting the hypothesis of recent inflow of gas at these sites.
The sample of GRB/SN hosts can then be used to study recent gas inflow. 
Our result of a very weak molecular deficiency (if any) implies that either the SFRs of GRB/SN hosts are not significantly enhanced by such inflow, or that atomic gas is efficiently converted into the molecular phase, so that SFR and {\mhtwo} increase hand in hand.

However, if molecular deficiency is confirmed with a larger sample of GRB hosts, then this will be consistent with a scenario in which their SFRs are enhanced by a recent inflow of atomic gas that did not have time to convert into the molecular phase. 
Moreover, a low molecular gas content would be consistent with star formation fuelled directly by atomic gas \citep{michalowski15hi}.

Two other issues need to be pointed out. First, most of our {\mhtwo} estimates are based on the CO(2-1) line or higher transitions. In order to calculate molecular gas masses, we converted these line luminosities into those of the CO(1-0) line assuming a conservatively low Milky Way {\lpcotwo/\lpcoone} ratio, giving  conservatively high {\mhtwo}. If however the gas in GRB hosts is even less excited than the Milky Way, then the real 2-1/1-0 ratio ratio is even lower, and our assumption would result in too low {\mhtwo}. This is unlikely, however, because GRB hosts are usually found to have a high SFR given their stellar masses    \citep{castroceron06,castroceron10,savaglio09,thone09}, which likely leads to high excitations \citep[see][]{michalowski16} and high {\lpcotwo/\lpcoone} ratios in turn. If this is the case generally, then our {\mhtwo} are overestimated, and the difference between GRB hosts and other galaxies is stronger than suggested by our analysis. 
In particular, if we were to adopt the SMG or M82 2-1/1-0 ratios, then the molecular gas masses of GRB hosts would be $1.7$--$2.0$ times lower \citep[table 2 of][]{carilli13}, and the difference between the GRB hosts and other galaxies would be statistically significant.
This can be tested with sensitive observations of other CO transitions (especially 1-0).

Second, our tentative molecular deficiency could result from the assumption of too low $\alpha_{\rm CO}$. We did take into account the variation of $\alpha_{\rm CO}$ with metallicity (Fig.~\ref{fig:sfrmh2metal}), but it is possible that other properties (e.g. gas density or turbulence) lead to high $\alpha_{\rm CO}$ and result in weak CO emission. 
On the other hand, if the correct $\alpha_{\rm CO}$ for GRB hosts is closer to the low value measured for starbursts \citep{bolatto13}, then the real molecular masses of GRB hosts are approximately five times lower than we measure and the molecular deficiency is statistically significant.
This aspect is much more difficult to investigate (also for non-GRB galaxies), because there is no robust way of measuring $\alpha_{\rm CO}$, especially in non-standard environments.

We also stress that it is important to investigate the molecular gas properties with high-resolution observations. If a molecular deficiency is found locally close to the GRB positions, then this will be consistent with star formation fuelled directly by atomic gas. In such a scenario, 
we are not able to capture this effect using the existing CO data with low spatial resolution, 
as the hosts on average are not significantly molecule poor.

This analysis can be improved by investigating a larger sample of GRB hosts, and possibly with deeper observations that allowing probing well below the average molecular gas depletion time of other star-forming galaxies.
Moreover, the caveat of our sample is that it is heterogenous, including low-$z$ hosts and highly star-forming hosts at higher redshifts \citep{hunt11,hunt14,svensson12,perley15}. This demonstrates the need of obtaining CO data for a larger sample of homogeneously selected GRB hosts. This is likely possible only with ALMA, because we have targeted nearby and bright hosts with CO emission that is potentially easier to detect. ALMA will be able to detect fainter targets 
and thus will enable studies of
a larger and unbiased sample.

\section{Conclusions}
\label{sec:conclusion}

We observed the CO(2-1) line for 7 GRB hosts, 
obtaining detections for 3 GRB hosts (980425, 080207, and 111005A) and upper limits for the remaining 4 (031203, 060505, 060814, and 100316D). 
In our entire sample of 12 CO-observed GRB hosts, including objects from the literature, 3 are clearly deficient in molecular gas, even taking into account their metallicity (980425, 060814, and 080517).
Four others are close to the best-fit line for other star-forming galaxies in the SFR-{\mhtwo} plot (051022, 060505, 080207, and 100316D). 
One host is clearly molecule rich (111005A). 
Finally, for 4 GRB hosts, the data are not deep enough to judge whether they are molecule deficient (000418, 030329, 031203, and 090423).
The median value of the molecular gas depletion time, $\mhtwo/\mbox{SFR}$, of GRB hosts is $\sim0.3$\,dex below that of other star-forming galaxies, but this result has low statistical significance. A Kolmogorov-Smirnov test performed on $\mhtwo/\mbox{SFR}$ shows only $\sim2\sigma$ difference between GRB hosts and other galaxies. This difference can partially be explained by metallicity effects, 
since the significance decreases to $\sim1\sigma$ for $\mhtwo/\mbox{SFR}$ versus~metallicity.

We found that any molecular gas deficiency of GRB hosts has low statistical significance and that it can be attributed to their
lower metallicities; and thus 
the sample of GRB hosts has consistent molecular properties to other galaxies and can be treated as representative of star-forming galaxies.
However, the molecular gas deficiency can be strong for GRB hosts if they exhibit higher excitations and/or a lower CO-to-{\htwo} conversion factor than we assume, which would lead to lower molecular gas masses than we derive.
Given the concentration of atomic gas recently found close to GRB and SN sites, indicating recent gas inflow, our results about the weak molecular deficiency imply that such inflow does not enhance the SFRs significantly, or that atomic gas converts efficiently into the molecular phase, which fuels star formation. Only if the analysis of a larger GRB host sample reveals molecular deficiency (especially close to the GRB position) would this support the hypothesis of star formation fuelled directly by atomic gas.

\begin{acknowledgements}

We thank Joanna Baradziej and our referee for help with improving this paper, Per Bergman, Carlos De Breuck, Palle M\o ller and Katharina Immer  for help with the APEX observations, and Claudia Marka for help with IRAM30m observations.

M.J.M.~acknowledges the support of 
the National Science Centre, Poland through the POLONEZ grant 2015/19/P/ST9/04010;
and the UK Science and Technology Facilities Council;
this project has received funding from the European Union's Horizon 2020 research and innovation programme under the Marie Sk{\l}odowska-Curie grant agreement No. 665778.
A.K.~acknowledges support from the Polish National Science Center grants 2014/15/B/ST9/02111and 2016/21/D/ST9/01098.
J.R.R. acknowledges the support from project ESP2015-65597-C4-1-R (MINECO/FEDER).
A.J.C.T.~acknowledges support from the Spanish Ministry Project AYA2015-71718-R.
J.H. was supported by a VILLUM FONDEN Investigator grant (project number 16599).
L.K.H.~acknowledges funding from the INAF PRIN-SKA program 1.05.01.88.04.
M.R.K.~acknowledges support from the Australian government through the Australian Research Council's Discovery Projects funding scheme (project DP160100695).

Based on observations collected at the European Organisation for Astronomical Research in the Southern Hemisphere under ESO programmes 096.D-0280(A), 096.F-9302(A), and 097.F-9308(A). 
This publication is based on data acquired with the Atacama Pathfinder Experiment (APEX). APEX is a collaboration between the Max-Planck-Institut fur Radioastronomie, the European Southern Observatory, and the Onsala Space Observatory.  
This work is based on observations carried out under project number 172-16 with the IRAM 30m telescope. IRAM is supported by INSU/CNRS (France), MPG (Germany) and IGN (Spain). 
This research has made use of 
the GHostS database (\urltt{http://www.grbhosts.org}), which is partly funded by Spitzer/NASA grant RSA Agreement No. 1287913; 
the NASA/IPAC Extragalactic Database (NED) which is operated by the Jet Propulsion Laboratory, California Institute of Technology, under contract with the National Aeronautics and Space Administration;
SAOImage DS9, developed by Smithsonian Astrophysical Observatory \citep{ds9};
and the NASA's Astrophysics Data System Bibliographic Services.

\end{acknowledgements}



\appendix
\section{Additional table}

\newcommand{\GiveRef}[1]{\citetalias{#1}: \citet{#1}}

\defcitealias{tinney98}{1}
\defcitealias{bloom03b}{2}
\defcitealias{greiner03gcn}{3}
\defcitealias{hjorthnature}{4}
\defcitealias{prochaska04}{5}
\defcitealias{castrotirado07}{6}
\defcitealias{ofek06gcn}{7}
\defcitealias{hjorth12}{8}
\defcitealias{stanway15b}{9}
\defcitealias{tanvir09}{10}
\defcitealias{salvaterra09}{11}
\defcitealias{vergani10gcn}{12}
\defcitealias{starling11}{13}
\defcitealias{michalowski18grb}{14}
\defcitealias{levan11gcn}{15}
\defcitealias{pignata09cbet}{16}
\defcitealias{michalowski14}{17}
\defcitealias{perley17b}{18}
\defcitealias{michalowski12grb}{19}
\defcitealias{watson11}{20}
\defcitealias{hunt14}{21}
\defcitealias{michalowski15hi}{22}
\defcitealias{perley15}{23}
\defcitealias{walter12}{24}
\defcitealias{tanga18}{25}
\defcitealias{michalowski18}{26}
\defcitealias{sollerman05}{27}
\defcitealias{christensen08}{28}
\defcitealias{svensson10}{29}
\defcitealias{levesque10c}{30}
\defcitealias{thone08}{31}
\defcitealias{kruhler15}{32}
\defcitealias{levesque11}{33}

\begin{table*}
\scriptsize
\centering
\caption{Properties of our sample of GRB hosts.\label{tab:sample}}
\begin{tabular}{llccccc}
\hline
GRB & $z_{\rm opt}$ & Ref & SFR & Ref & $\metoh$ & Ref\\
    &              & & ($M_\odot\mbox{ yr}^{-1}$) & &   \\
\hline
980425 & 0.0085 & \citetalias{tinney98} & $0.26$ & \citetalias{michalowski14} & 8.60 & \citetalias{sollerman05} \\
980425\_WR & 0.0085 & \citetalias{tinney98} & $0.02$ & \citetalias{michalowski14} & 8.16 & \citetalias{christensen08} \\
000418 & 1.1181 & \citetalias{bloom03b} & $<77$ & \citetalias{perley17b} & 8.43 & \citetalias{svensson10} \\
030329 & 0.1685 & \citetalias{greiner03gcn},\citetalias{hjorthnature} & $<17$ & \citetalias{michalowski12grb} & 8.13 & \citetalias{levesque10c} \\
031203 & 0.1050 & \citetalias{prochaska04} & $2.8$ & \citetalias{watson11} & 8.27 & \citetalias{levesque10c} \\
051022 & 0.809 & \citetalias{castrotirado07} & $17.9$ & \citetalias{hunt14} & 8.62 & \citetalias{levesque10c} \\
060505 & 0.0889 & \citetalias{ofek06gcn} & $0.69$ & \citetalias{michalowski15hi} & 8.30 & \citetalias{thone08} \\
060814 & 1.9229 & \citetalias{hjorth12} & $256$ & \citetalias{perley15} & 8.38 & \citetalias{kruhler15} \\
080207 & 2.0858 & \citetalias{hjorth12} & $170$ & \citetalias{hunt14} & 8.74 & \citetalias{kruhler15} \\
080517 &        0.089 & \citetalias{stanway15b} & $7.6$ & \citetalias{stanway15b} & 8.66 & \citetalias{stanway15b} \\
090423 & 8.23     & \citetalias{tanvir09}, \citetalias{salvaterra09} & $<39$ & \citetalias{walter12} & - & - \\
100316D & 0.0591 & \citetalias{vergani10gcn},\citetalias{starling11} & $1.73$ & \citetalias{michalowski15hi} & 8.30 & \citetalias{levesque11} \\
111005A & 0.01326 & \citetalias{michalowski18grb},\citetalias{levan11gcn} & $0.42$ & \citetalias{michalowski18grb}  & 8.50 & \citetalias{tanga18} \\
111005A\_CENT & 0.01326 & \citetalias{michalowski18grb},\citetalias{levan11gcn} & $0.26$ & \citetalias{tanga18}  & 8.56 & \citetalias{tanga18} \\
111005A\_NW & 0.01326 & \citetalias{michalowski18grb},\citetalias{levan11gcn} & $0.06$ & \citetalias{tanga18}  & 8.49 & \citetalias{tanga18} \\
111005A\_SE & 0.01326 & \citetalias{michalowski18grb},\citetalias{levan11gcn} & $0.09$ & \citetalias{tanga18}  & 8.43 & \citetalias{tanga18} \\
SN\,2009bb & 0.009877 & \citetalias{pignata09cbet} & $5.21$ & \citetalias{michalowski18}  & 8.66 & \citetalias{michalowski18} \\
\hline
\end{tabular}

References: \GiveRef{tinney98},  
\GiveRef{bloom03b},
\GiveRef{greiner03gcn},
\GiveRef{hjorthnature},
\GiveRef{prochaska04}, 
\GiveRef{castrotirado07},
\GiveRef{ofek06gcn},  
\GiveRef{hjorth12},
\GiveRef{stanway15b},
\GiveRef{tanvir09},
\GiveRef{salvaterra09},
\GiveRef{vergani10gcn},  
\GiveRef{starling11}, 
\GiveRef{michalowski18grb}, 
 \GiveRef{levan11gcn},  
 \GiveRef{pignata09cbet},
 \GiveRef{michalowski14}, 
 \GiveRef{perley17b},
\GiveRef{michalowski12grb},
\GiveRef{watson11},
\GiveRef{hunt14},
\GiveRef{michalowski15hi},
 \GiveRef{perley15},
\GiveRef{walter12},
\GiveRef{tanga18},
\GiveRef{michalowski18},
\GiveRef{sollerman05},  
\GiveRef{christensen08},
\GiveRef{svensson10},
\GiveRef{levesque10c}, 
\GiveRef{thone08},  
\GiveRef{kruhler15},
 \GiveRef{levesque11}, 
\end{table*}

\end{document}